\documentclass[preprint,12pt]{elsarticle}

\usepackage{newtxtext,newtxmath}
\usepackage[scaled=0.9]{inconsolata}

\usepackage[T1]{fontenc}
\usepackage[utf8]{inputenc}

\usepackage{amsmath}
\usepackage{mathtools}
\usepackage{bm}
\usepackage{mathrsfs}

\usepackage{graphicx}
\usepackage{float}
\usepackage{subcaption}
\usepackage{multirow,booktabs}

\usepackage{algorithm,algpseudocode}

\usepackage[x11names]{xcolor}

\usepackage{natbib}
\usepackage{hyperref}

\usepackage{lineno}

\usepackage[margin=2cm]{geometry}

\makeatletter
\algrenewcommand\ALG@beginalgorithmic{\ttfamily}
\makeatother

\makeatletter
\def\ps@pprintTitle{%
  \let\@oddhead\@empty
  \let\@evenhead\@empty
  \let\@oddfoot\@empty
  \let\@evenfoot\@oddfoot
}
\makeatother

\begin{document}
\begin{frontmatter}

\title{Probabilistic multivariate statistical process control via kernel parameter uncertainty propagation}


\author[1]{Zina-Sabrina Duma\corref{cor1}}
\cortext[cor1]{Corresponding author}
\ead{Zina-Sabrina.Duma@lut.fi}
\author[1]{Victoria Jorry}
\author[1]{Ayesha Sarfraz}
\author[1]{Maria Paola di Crosta}
\author[1]{Tuomas Sihvonen}
\author[1]{Lassi Roininen}
\author[1]{Satu-Pia Reinikainen}

\address[1]{LUT University, Yliopistonkatu 34, Lappeenranta 53850, Finland}

\begin{abstract}
Kernel-based multivariate statistical process control (K-MSPC) extends classical monitoring to nonlinear industrial processes. Its performance depends critically on kernel parameters such as lengthscales and variance terms. In current practice these parameters are typically selected by heuristics or deterministic optimisation, and then treated as fixed, despite being inferred from finite and noisy data. This can lead to overconfident control limits and unstable alarm behaviour when the kernel choice is uncertain. This work proposes a probabilistic K-MSPC framework that quantifies and propagates kernel parameter uncertainty to the monitoring statistics. The approach follows a two-stage workflow: (i) deterministic kernel calibration using supervised or unsupervised models, and (ii) Bayesian inference of kernel parameters via Markov chain Monte Carlo. Posterior samples are propagated through kernel Principal Component Analysis to produce probabilistic $T^2$ and squarred prediction error control charts, together with uncertainty-aware contribution plots. The framework is evaluated on the Tennessee Eastman Process benchmark. Results show that posterior-mean monitoring often improves fault detection compared to deterministic prior-mean charts for the squared exponential kernel, while credible bands remain narrow in-control and widen under faults, reflecting amplified epistemic uncertainty in abnormal regimes. The automatic relevance determination kernel reduces posterior uncertainty and yields performance close to the deterministic baseline, whereas unsupervised calibration produces wider posterior bands but still robust fault detection.
\end{abstract}

\begin{keyword}
Kernel PCA \sep Bayesian inference \sep MCMC \sep Uncertainty quantification \sep Tennessee Eastman Process
\end{keyword}

\end{frontmatter}

\section{Introduction}\label{sc:Intro}
Modern industrial plants are increasingly instrumented with high-frequency, high-dimensional measurements, creating a strong need for reliable data-driven monitoring methods that can detect abnormal operating conditions early and support diagnosis \cite{Panaretos2005}. Multivariate statistical process control (MSPC) remains a cornerstone in this space, where latent-variable models such as principal component analysis (PCA) translate multivariate measurements into low-dimensional monitoring statistics and corresponding control charts, typically Hotelling's $T^2$ and squared prediction error (SPE) indicators, for fault detection and interpretation \citep{KourtiMacGregor1995,NomikosMacGregor1995}. 
In contemporary manufacturing settings, MSPC is frequently embedded within broader “data science for operations” and Industry~4.0 initiatives, where statistical process control and related monitoring practices are combined with modern analytics workflows \citep{Pozzi2024}. 

While classical MSPC based on PCA  has proven effective for many industrial applications, its linear structure limits its ability to capture complex nonlinear relationships commonly encountered in modern processes \cite{Scholkopf1998KernelPCA}. To address this limitation, nonlinear extensions such as Kernel PCA (K-PCA) are introduced by mapping data into a high-dimensional reproducing kernel Hilbert space (RKHS), where linear projections correspond to nonlinear structures in the original space \citep{Scholkopf1998KernelPCA}. Kernel-based MSPC (K-MSPC) has since demonstrated improved fault detection capabilities in processes exhibiting nonlinear dynamics, multimodality, and complex interactions \cite{liu2023multivariate, lee2020monitoring, zhou2020multi}.

However, this increased modelling flexibility introduces an additional layer of complexity: the selection of kernel parameters \cite{simmini2021self}, such as the length scale and variance parameters in radial basis function kernels. These parameters directly influence the geometry of the feature space, the variance captured by principal components, and ultimately the sensitivity of $T^2$ and SPE monitoring statistics \cite{tan2020}. In most existing K-MSPC studies, kernel parameters are selected using heuristic rules (e.g., median distance criteria), cross-validation, or deterministic optimisation procedures \cite{Alam2014HyperparameterSI, DUMA2026103679}. Once chosen, they are treated as fixed quantities, and subsequent monitoring proceeds under the implicit assumption that the selected kernel represents the true data-generating structure. Optimisation procedures such as gradient-based methods, genetic algorithms, or heuristic selection rules yield point estimates that are subsequently treated as fixed and certain \cite{pani2022non}. However, kernel parameters are not directly observable quantities; they are inferred from finite and often noisy training data. As such, they are subject to epistemic uncertainty arising from limited sample size, model misspecification, and optimisation landscape nonconvexity.

Ignoring this uncertainty has important implications for statistical process monitoring. The geometry of the kernel feature space — and therefore the eigenstructure of the K-PCA model — depends sensitively on the chosen parameters. Small variations in the lengthscale, for instance, can significantly alter the variance distribution between the retained principal subspace and the residual space \cite{DUMA2026103679}. Consequently, control limits derived from deterministic kernel estimates may be overconfident, potentially underestimating variability under normal conditions or mischaracterising fault persistence. This can manifest as unstable alarm behaviour, inflated false alarms, or premature return-to-normal indications in control charts. In other words, while K-MSPC extends linear MSPC to nonlinear settings, it inherits a hidden layer of uncertainty that remains unquantified in standard practice.

To address this limitation, the objective of this work is to develop a probabilistic kernel-based multivariate statistical process control framework that explicitly quantifies and propagates uncertainty in kernel parameters to monitoring statistics. Rather than treating kernel parameters as fixed point estimates, we model them as random variables and infer their posterior distributions using Bayesian techniques. The resulting uncertainty is then propagated through the Kernel PCA model to obtain probabilistic control charts for both the $T^2$ and SPE statistics, as well as for their corresponding variable contribution diagnostics. By embedding kernel calibration within a Bayesian inference framework, the proposed approach enables monitoring decisions that reflect epistemic uncertainty in model structure. This leads to control charts that not only signal faults but also quantify confidence in detection and attribution, thereby enhancing robustness, interpretability, and reliability in nonlinear process monitoring.

\subsection{Our contribution}

This work is a direct continuation of Duma et al. (2026) \cite{DUMA2026103679}, which addressed deterministic kernel calibration for K-MSPC through supervised and procedural optimisation. The present paper addresses the complementary question left open there: how uncertainty in the calibrated kernel parameters propagates to monitoring statistics, control limits, and contribution diagnostics.

The main contribution of this paper is: 
\begin{itemize}
    \item Process monitoring using control charts is more accurate under uncertain and nonlinear conditions, thereby enhancing industrial quality yield and productivity.
\end{itemize}
This contribution is achieved by the following sub-contributions: 
\begin{enumerate}[I.]
    \item  We propose a two-stage probabilistic calibration framework for K-MSPC. In the first stage, kernel parameters are estimated deterministically using both supervised and unsupervised strategies. The supervised route employs intermediary classification or regression models (Gaussian process classification, GPC \cite{nickisch2008approximations}, and Kernel PCR \cite{DUMA2026103679}) optimised via gradient-based and derivative-free methods, while the unsupervised route evaluates a suite of kernel-selection criteria tailored to healthy-operation data. This stage provides informed prior means for subsequent Bayesian inference.
    \item We introduce Bayesian sampling of kernel parameters using Markov chain Monte Carlo (MCMC) methods, including adaptive Metropolis (AM) \cite{haario2001adaptive}, delayed rejection adaptive Metropolis (DRAM) \cite{haario2006dram}, Hamiltonian Monte Carlo (HMC) \cite{betancourt2017conceptual}, and the No-U-Turn Sampler (NUTS) \cite{HoffmanGelman2014NUTS}. This enables posterior inference over kernel lengthscales and variance parameters rather than reliance on single-point estimates.
    \item We formulate probabilistic $T^2$ and SPE control charts by propagating posterior uncertainty through the Kernel PCA monitoring framework. In addition to probabilistic monitoring statistics, we derive uncertainty-aware contribution plots, allowing variable attribution to be expressed together with credible intervals.
\end{enumerate}

The proposed framework differs conceptually from three related research directions. First, while Bayesian inference of kernel parameters is well established in Gaussian process modelling, existing work focuses on predictive uncertainty of the GP output rather than on how parameter uncertainty alters downstream monitoring structure. Here, the posterior distribution is not an end in itself but a mechanism for reshaping the K-PCA eigenspace and the associated monitoring statistics. Second, unlike conventional K-MSPC approaches that rely on a single optimised kernel configuration, the proposed formulation replaces fixed-kernel monitoring with a distribution over monitoring models, thereby exposing the sensitivity of control charts to kernel calibration. Third, in contrast to uncertainty bands derived from bootstrap resampling or measurement noise around a fixed model, the uncertainty quantified in this work is structural and epistemic: it originates from kernel calibration itself and propagates through both detection statistics and diagnostic contributions. This perspective reframes kernel selection not as a preprocessing step but as a source of model uncertainty that directly affects alarm stability, fault persistence, and interpretability.

The framework is evaluated using the Tennessee Eastman process (TEP) \cite{chiang2001tennessee} benchmark, demonstrating how parameter uncertainty affects fault detection performance, control-limit stability, and interpretability under both supervised and unsupervised calibration schemes.

\subsection{Outline of the paper}

This paper is organized as follows: the proposed framework is described in Section \ref{sc:MatMet}, and additional notations are expanded in \ref{app:notation}. The intermediary results on prior mean learning and the  probabilistic control charts are presented and discussed in Section \ref{sc:results} and expanded in \ref{app:detvsprob} and \ref{app:KPCRResults}.

\section{Materials and methods}\label{sc:MatMet}

This paper extends the deterministic kernel-calibration framework introduced in Duma \textit{et al.}, 2026 \cite{DUMA2026103679} to a probabilistic setting. In the earlier work, kernel parameters were learned procedurally for K-MSPC using a supervised K-PCR-based intermediary objective, and the TEP benchmark was used for evaluation. In the present study, the focus shifts from deterministic point estimation to posterior uncertainty quantification and propagation. We therefore adopt the notation and deterministic K-PCA/K-MSPC framework \cite{DUMA2026103679} where possible, and introduce here only the additional elements required for Bayesian inference, probabilistic control charts, Gaussian-process-based calibration, and the unsupervised calibration route.

\subsection{Overview of the proposed probabilistic framework}\label{ssc:overview}

\begin{figure}[H]
    \centering
    \includegraphics[width=\linewidth]{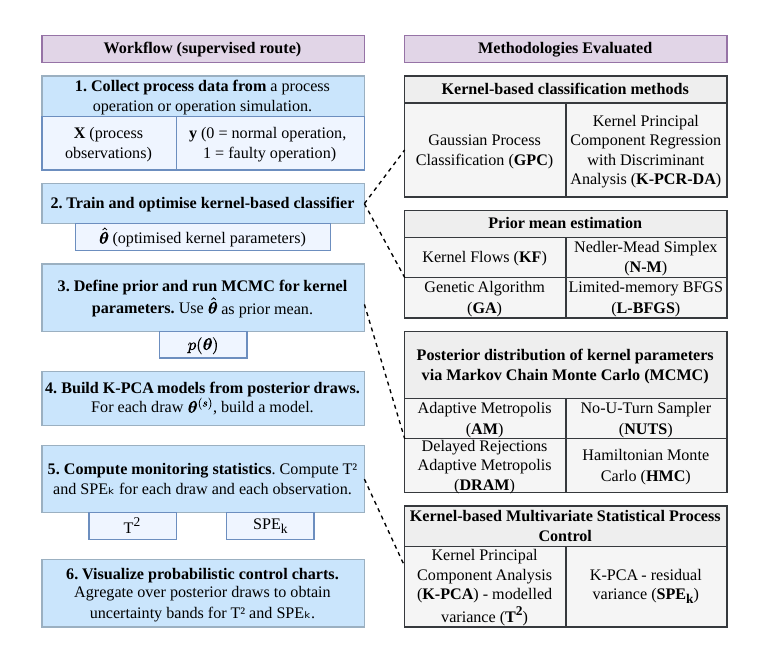}
    \caption{Workflow for the generation of probabilistic control charts. Deterministic kernel calibration provides prior means, Bayesian sampling yields posterior draws of kernel parameters, and posterior propagation through K-PCA produces probabilistic monitoring statistics and contribution diagnostics. Here, is $\mathbf{X}$ a matrix of process observations (timeline) as rows, and process variables as columns, and \textbf{y} is the process state (either '0' if the process is in normal operation or '1' if the process is faulty). }
    \label{fig:workflow}
\end{figure}

The proposed methodology consists of three connected stages. The workflow is illustrated in Figure~\ref{fig:workflow}. First, kernel parameters are calibrated deterministically in order to obtain an informative reference point for inference. In the supervised route, this calibration is performed using labelled healthy/faulty data through either GPC or K-PCR-based discrimination. In the unsupervised route, only healthy-operation data are used and kernel parameters are selected using dataset-level structural criteria.

Second, the deterministic estimate is treated as the prior mean in a Bayesian formulation for the kernel parameters. A posterior distribution is then inferred with MCMC, thereby quantifying epistemic uncertainty in the kernel lengthscale and variance parameters.Third, posterior draws of the kernel parameters are propagated through the K-PCA monitoring model. Each draw induces a corresponding K-PCA representation and therefore a corresponding pair of monitoring statistics, $T^2$ and SPE, together with variable contribution diagnostics. Aggregating over posterior draws yields probabilistic control charts and uncertainty-aware contribution plots.

The resulting framework may be interpreted as a probabilistic extension of deterministic K-MSPC: instead of a single calibrated monitoring model, one obtains a posterior distribution over monitoring models induced by uncertainty in the kernel parameters. 

\subsection{Bayesian formulation for kernel parameter uncertainty}\label{ssc:bayes}

Let the calibration dataset be denoted by $\mathcal{D}=\{(\mathbf{x}_i,y_i)\}_{i=1}^n$, where $\mathbf{x}_i \in \mathbb{R}^p$ represents the $i$-th observation of an assumed gaussian multivariate distribution \textbf{X} and $y_i\in\{0,1\}$ denotes an associated label indicating healthy or faulty operation when such labels are available. 
We assume that the mapping between $\mathbf{x}_i$ and $y_i$ depends on a parametric model, with unknown parameters $\boldsymbol \theta \in \mathbb R^{m}$.
Then, we are interested in estimating the posterior distribution of $\boldsymbol \theta$ given the dataset $\mathcal{D}$. 
We work in the Bayesian framework, and the solution is given as a posterior distribution 
\begin{equation}
p(\boldsymbol{\theta}\mid\mathcal{D})
\propto
p(\mathcal{D}\mid\boldsymbol{\theta})\,p(\boldsymbol{\theta}),
\end{equation}
where $p(\mathcal{D}\mid\boldsymbol{\theta})$ is the likelihood associated with the chosen intermediary model, and $p(\boldsymbol{\theta})$ is the prior distribution. We denote $\boldsymbol{\theta}$ as the kernel  vector. 

We consider two kernel families in this study, with the central modelling assumption that the kernel parameters should not be treated as fixed after calibration. 
Instead, they are regarded as uncertain quantities inferred from finite process data.  First we consider the isotropic squared exponential (SE), which has the form
\begin{equation}
k_{\mathrm{SE}}(\mathbf{x}_i,\mathbf{x}_j)
=
s_f^2
\exp\!\left(
-\frac{1}{2}\frac{\|\mathbf{x}_i-\mathbf{x}_j\|^2}{\ell^2}
\right),
\end{equation}
where \textbf{$\mathbf{x}_i$} is the $i$-th sample and $\mathbf{x}_j$ is the $j$-th sample. For the SE kernel, the kernel vector is 
\[
\boldsymbol{\theta}_{\mathrm{SE}}
=
\big[\ell, s_f, s_n\big],
\]
where $\ell > 0$ is the global lengthscale, $s_f^2$ is the signal variance, and $s_n^2$ is the noise variance used in the intermediary likelihood. 
For the Automatic Relevance Determination (ARD) formulation,
\begin{equation}
k_{\mathrm{ARD}}(\mathbf{x}_i,\mathbf{x}_j)
=
s_f^2
\exp\!\left(
-\frac{1}{2}\sum_{d=1}^{p}
\frac{(x_{i,d}-x_{j,d})^2}{\ell_d^2}
\right),
\end{equation}
where $x_{i,d}$ is the $i$-th sample value in the $d$-th variable, and $x_{j,d}$ is the $j$-th sample value in the $d$-th variable. The kernel vector takes the form
\[
\boldsymbol{\theta}_{\mathrm{ARD}}
=
\big[\ell_1,\ldots,\ell_p,s_f,s_n\big],
\]
where each variable has its own characteristic lengthscale.

Choosing the prior distribution is crucial to the problem, as it is the only tuneable object after the kernel has been fixed. 
We choose weakly informative log-Gaussian models for the parameters, and for the prior mean, we  use a deterministic estimate $\widehat{\boldsymbol{\theta}}$, obtained from either supervised or unsupervised calibration. 
The deterministic kernel estimation is further presented in Section \ref{ssc:deterministicCalibration}. The priors are then 
\[
\log \ell \sim \mathcal{N}({\log \widehat \ell},\sigma^2_{\log \ell}), \qquad
\log s_f \sim \mathcal{N}({\log \widehat s_f},\sigma^2_{\log s_f}), \qquad
\log s_n \sim \mathcal{N}({\log \widehat s_n},\sigma^2_{\log s_n}).
\]
We use analogous priors for the ARD lengthscales $\log(\ell_d)$. This parameterisation ensures positivity of the kernel quantities while allowing moderate posterior dispersion around the deterministic solution.

Two supervised and one unsupervised intermediary likelihood constructions are considered: 

\begin{itemize}
    \item \textbf{In the first supervised route}, kernel parameters are inferred through Gaussian process classification. A Gaussian process prior is placed on the latent function,
\[
f(\mathbf{x}) \sim \mathcal{GP}\!\big(0,k_{\boldsymbol{\theta}}(\mathbf{x},\mathbf{x}')\big),
\]
where $k_{\boldsymbol{\theta}}$ is either $k_{SE}$ or $k_{ARD}$. For computational convenience, the likelihood entering the posterior is written through the corresponding Gaussian-process marginal form,


\begin{equation}
    p(\mathcal D \mid \boldsymbol{\theta}) = \frac{1}{\sqrt{(2\pi)^n\vert \mathbf{C}_{\boldsymbol{\theta}}\vert}} \exp\left( -\frac{1}{2} \mathbf{y}^\top\mathbf{C}_{\boldsymbol{\theta}}^{-1}\mathbf{y}\right),
\qquad
\mathbf{C}_{\boldsymbol{\theta}}=\mathbf{K}_{\boldsymbol{\theta}}+s_n^2\mathbf{I}.
\end{equation}
where $\mathbf{K}_{\boldsymbol{\theta}}$ is the kernel matrix induced by the current parameters. The associated log-likelihood is
\[
\log p(\mathcal D \mid \boldsymbol{\theta})
=
-\tfrac12\mathbf{y}^{\top}\mathbf{C}_{\boldsymbol{\theta}}^{-1}\mathbf{y}
-\tfrac12\log|\mathbf{C}_{\boldsymbol{\theta}}|
-\tfrac{n}{2}\log(2\pi)
\]

    \item \textbf{In the second supervised route}, kernel parameters are inferred through K-PCR-based discrimination, extending the deterministic K-PCR calibration framework of Duma \textit{et al.}, 2026 \cite{DUMA2026103679} into a Bayesian setting. Let $\mathbf{T}_{\boldsymbol{\theta}}\in\mathbb{R}^{n\times r}$ denote the K-PCA score matrix induced by kernel parameters $\boldsymbol{\theta}$, and let $\hat{\mathbf{y}}=\mathbf{T}_{\boldsymbol{\theta}}\boldsymbol{\beta}$ denote a linear regression model in score space. Assuming Gaussian residuals,
\[
\boldsymbol{\varepsilon}
=
\mathbf{y}-\mathbf{T}_{\boldsymbol{\theta}}\boldsymbol{\beta}
\sim
\mathcal{N}(\mathbf{0},\sigma^2\mathbf{I}),
\]
the likelihood becomes
\[
p(\mathcal D \mid \boldsymbol{\theta})
=
\mathcal{N}\!\big(\mathbf{y}\mid\mathbf{T}_{\boldsymbol{\theta}}\boldsymbol{\beta},\,\sigma^2\mathbf{I}\big)
\propto
\exp\!\Big(
-\tfrac{1}{2\sigma^2}
\|\mathbf{y}-\mathbf{T}_{\boldsymbol{\theta}}\boldsymbol{\beta}\|^2
\Big).
\]

    \item \textbf{In the unsupervised route}, no observed fault labels are initially available. Let $\mathcal{D}=\{(\mathbf{x}_i,\tilde y_i)\}_{i=1}^n$, where $\tilde y_i\in\{0,1\}$ denotes the chart-induced pseudo-label assigned after deterministic calibration on healthy-operation data. Kernel parameters are first calibrated deterministically from healthy-operation data only, after which a provisional K-PCA monitoring model is constructed and observations are assigned induced labels $\tilde y_i\in\{0,1\}$ according to whether their monitoring statistics remain within or exceed the chart limits. These induced labels are then used in the Bayesian stage through the same Gaussian-process marginal surrogate likelihood as in the first supervised route, namely
\begin{equation}
    \begin{split}
p(\mathcal D \mid \boldsymbol{\theta}) &= \frac{1}{\sqrt{(2\pi)^n\vert \mathbf{C}_{\boldsymbol{\theta}}\vert}} \exp\left( -\frac{1}{2} \widetilde{\mathbf{y}}^\top\mathbf{C}_{\boldsymbol{\theta}}^{-1}\widetilde{\mathbf{y}}\right), \\
\log p(\mathcal D \mid \boldsymbol{\theta})
&=
-\tfrac12\tilde{\mathbf{y}}^{\top}\mathbf{C}_{\boldsymbol{\theta}}^{-1}\tilde{\mathbf{y}}
-\tfrac12\log|\mathbf{C}_{\boldsymbol{\theta}}|
-\tfrac{n}{2}\log(2\pi).
    \end{split}
\end{equation}
The resulting posterior should therefore be interpreted as conditional on chart-induced pseudo-labels rather than externally observed fault classes.
\end{itemize}

In all three intermediary constructions, the posterior captures uncertainty in the kernel parameters induced by the calibration data and the chosen surrogate model. Posterior sampling is performed with four Markov chain Monte Carlo strategies: Adaptive Metropolis (AM) \cite{haario2001adaptive}, Delayed Rejection Adaptive Metropolis (DRAM) \cite{haario2006dram}, Hamiltonian Monte Carlo (HMC) \cite{betancourt2017conceptual}, and the No-U-Turn Sampler (NUTS) \cite{HoffmanGelman2014NUTS}. Sampling is carried out in log-parameter scale, with chains initialised at the deterministic estimate $\widehat{\boldsymbol{\theta}}$. After burn-in removal, posterior summaries are computed from the retained draws.

Convergence and sampling efficiency are assessed using trace plots, autocorrelation functions, effective sample size (ESS) \cite{kavianihamedani2024new}, and stability of posterior summaries across sampling windows. Agreement between different samplers is used as an additional qualitative robustness check.

\subsection{Probabilistic monitoring statistics and posterior propagation}\label{ssc:propagation}

The deterministic K-PCA/K-MSPC machinery follows Duma \textit{et al.}, 2026 \cite{DUMA2026103679}; only its probabilistic extension is introduced here. Let $\boldsymbol{\theta}^{(m)}$, $m=1,\ldots,M$, denote posterior draws from $p(\boldsymbol{\theta}\mid\mathcal{D})$. For each draw, a kernel matrix is constructed using either $k_{\mathrm{SE}}$ or $k_{\mathrm{ARD}}$, a K-PCA model is obtained, and the corresponding monitoring statistics are evaluated for each sample.

Thus, each posterior draw induces a draw-specific pair of monitoring statistics,
\[
T^{2\,(m)}(\mathbf{x}), \qquad \mathrm{SPE}^{(m)}(\mathbf{x}),
\]
as well as draw-specific contribution diagnostics,
\[
C_{d}^{T^2,(m)}(\mathbf{x}), \qquad C_{d}^{\mathrm{SPE},(m)}(\mathbf{x}).
\]

The posterior sample $\{T^{2\,(m)}(\mathbf{x})\}_{m=1}^{M}$ defines an empirical posterior distribution for the $T^2$ statistic at sample $\mathbf{x}$, and similarly for SPE and the contribution terms. Point summaries such as posterior means or medians can therefore be plotted as representative control charts, while credible intervals quantify the uncertainty induced by kernel parameter variability.

In this work, probabilistic control charts are constructed by summarising the posterior distribution of the monitoring statistics at each time point. In the same manner, uncertainty-aware contribution plots are obtained by summarising the posterior distribution of the contribution coefficients for each variable. This allows not only fault detection under uncertainty, but also uncertainty quantification of variable attribution.

An important conceptual point is that the uncertainty considered here is structural rather than measurement-level noise around a fixed model. Each posterior draw modifies the kernel geometry, which in turn changes the K-PCA eigenspace, the monitored subspace/residual decomposition, and consequently the values of $T^2$, SPE, and their contributions. The probabilistic control charts therefore reflect uncertainty in the monitoring model itself.

\subsection{Deterministic kernel calibration routes}\label{ssc:deterministicCalibration}

The role of deterministic calibration in the present study is to provide informative prior means for the Bayesian stage. Three calibration routes are considered: supervised calibration through GPC, supervised calibration through K-PCR, and unsupervised calibration from healthy-operation data only.

\subsubsection{Supervised calibration for pocesses with known fault information}

The first supervised route uses Gaussian process classification to distinguish healthy and faulty observations. Kernel parameters are estimated by deterministic optimisation of the intermediary objective with respect to $\boldsymbol{\theta}$. The optimisers considered are L-BFGS \cite{moritz2016linearly}, Nelder--Mead \cite{gao2012implementing}, Genetic Algorithm \cite{forrest1996genetic}, and Kernel Flows \cite{owhadi2019kernel}. Optimisation is performed in log-parameter scale to preserve positivity of the kernel quantities. This route is new relative to Duma \textit{et al.}, 2026 \cite{DUMA2026103679} and is included here because it provides an alternative supervised surrogate for kernel calibration before posterior inference. 

The second supervised route follows the deterministic K-PCR-based calibration philosophy introduced in Duma \textit{et al.}, 2026 \cite{DUMA2026103679}. In that framework, kernel parameters are selected through discrimination performance in the same truncated K-PCA representation used later for monitoring, making the calibration structurally aligned with K-MSPC. In the present paper, those deterministic K-PCR estimates are not an end point; instead, they are used as prior means for subsequent Bayesian sampling. Because this deterministic K-PCR framework has already been presented in detail, it is not re-derived here.

\subsubsection{Unsupervised calibration from healthy-operation data}

In many process-monitoring applications, labelled fault data are unavailable. To address this setting, an unsupervised calibration route is included in the present study. Here, kernel parameters are inferred exclusively from healthy-operation data before Bayesian posterior inference.

For the unsupervised route, calibration is restricted to the isotropic SE kernel with a single global lengthscale $\ell$. This choice favours parsimony and robustness, and avoids the severe ill-posedness that would arise when estimating many variable-specific lengthscales without supervision. Ten unsupervised selection rules are considered, each targeting a different structural property of the healthy-data kernel representation. These are summarised in Table~\ref{tab:unsupervised-kernel-methods}.

\begin{table}[H]
\caption{Unsupervised kernel-parameter tuning methods evaluated for deterministic K-MSPC.}
\label{tab:unsupervised-kernel-methods}
\centering
\small
\setlength{\tabcolsep}{4pt}
\renewcommand{\arraystretch}{1.15}

\begin{tabular}{p{4.0cm}|p{10cm}|p{2.3cm}}
\hline
\textbf{Method} & \textbf{Description} & \textbf{Inspired by} \\
\hline
\textbf{M1} Median all-pairs & RBF lengthscale set to the median of all pairwise distances computed from healthy operation data. & \cite{scholkopf2002learning,gretton2012kernel} \\ \hline
\textbf{M2} Median kNN
& Lengthscale set to the median of $k$-nearest neighbour distances, improving robustness of local similarity measures to outliers and nonstationarity. & \cite{zelnik2004self,hein2005intrinsic} \\ \hline
\textbf{M3} Target mean similarity & Kernel lengthscale chosen such that the average off-diagonal kernel similarity matches a predefined target value. & \cite{shawe2004kernel} \\ \hline
\textbf{M4} Effective rank target & Kernel tuned to achieve a desired effective rank (spectral entropy) of the centered Gram matrix, controlling model complexity. & \cite{roy2007kernel,bach2002kernel} \\ \hline
\textbf{M5} Split-half subspace & Lengthscale selected to maximise stability of K-PCA subspaces across random splits of healthy data. & \cite{meinshausen2010stability} \\ \hline
\textbf{M6} Quantile stability & Kernel selected to minimise variability of healthy T$^2$ and SPE control limits across resampled datasets. & \cite{nomikos1995monitoring} \\ \hline
\textbf{M7} One-class SVM proxy & Kernel tuned using one-class SVM criteria, such as support fraction and rejection stability, as a proxy objective. & \cite{scholkopf2001estimating,tax2004support} \\ \hline
\textbf{M8} MMD halves & Lengthscale chosen to minimise the maximum mean discrepancy between temporal halves of healthy operation data. & \cite{gretton2012kernel,borgwardt2006integrating} \\ \hline
\textbf{M9} KRR self-predict & Kernel ridge regression trained to reconstruct healthy data; lengthscale minimises cross-validated prediction error. & \cite{rifkin2003regularized,shawe2004kernel} \\ \hline
\textbf{M10} Alignment kNN graph & Kernel aligned with a $k$-nearest-neighbour graph to preserve local manifold structure of healthy data. & \cite{cortes2009shape,belkin2003laplacian} \\
\hline
\end{tabular}
\end{table}

\begin{figure}[H]
    \centering
    \includegraphics[width=\linewidth]{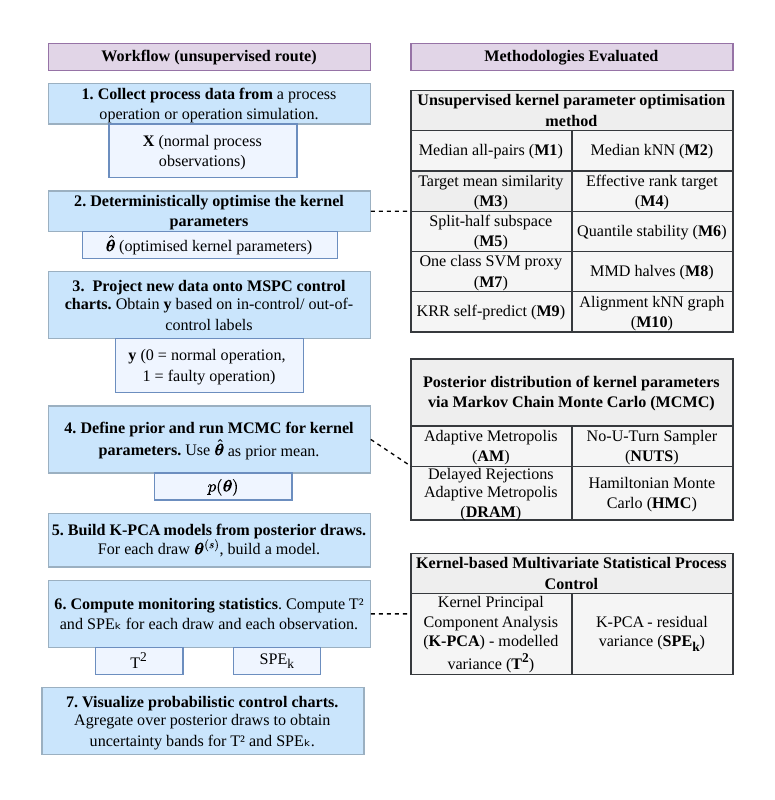}
    \caption{Workflow for generation of probabilistic control charts when no fault data are initially available. Deterministic unsupervised kernel calibration is followed by provisional chart-based label assignment and subsequent Bayesian posterior inference.}
    \label{fig:unsupervisedWorkflow}
\end{figure}

For this unsupervised route, the workflow differs from the supervised case only in the source of the prior mean. Normal-operation data are first used to select the deterministic kernel parameter. A K-PCA monitoring model is then fitted, control limits are computed, and new observations are provisionally mapped into in-control/out-of-control classes through their position relative to the chart limits. These induced labels are then used in the subsequent Bayesian sampling stage. The unsupervised workflow is illustrated in Figure~\ref{fig:unsupervisedWorkflow}.

The notation and deterministic K-PCA/K-MSPC framework follow Duma \textit{et al.}, 2026 \cite{DUMA2026103679}, including kernel centering, score computation, construction of $T^2$ and SPE control charts, and contribution diagnostics. To avoid repeating material from the earlier paper, these definitions are summarised in \textbf{\ref{app:notation}}. The present paper introduces only the additional notation associated with kernel parameter posteriors, MCMC sampling, and posterior propagation.

\subsection{Case study: Tennessee Eastman Process}\label{ssc:TEP}

The proposed framework is evaluated on the Tennessee Eastman Process benchmark, following the same general process setting as in Duma \textit{et al.}, 2026 \cite{DUMA2026103679}. The TEP is a widely used nonlinear industrial process simulation with interacting process and manipulated variables and 21 predefined fault scenarios. Detailed process and variable descriptions are available in \cite{downs1993plant}, \cite{chiang2001tennessee}, and \cite{DUMA2026103679}.

In the present work, healthy-operation data are used to construct the monitoring model and to support unsupervised calibration, while healthy/faulty labelled data are additionally used in the supervised intermediary calibration routes. Independent test partitions with fault onset during the monitoring trajectory are used to evaluate deterministic and probabilistic chart performance. This setup preserves comparability with the earlier deterministic study while allowing the new probabilistic framework to be assessed under both supervised and unsupervised calibration conditions.

\section{Results and discussion}\label{sc:results}

\subsection{Supervised calibration}

\subsubsection{Supervised intermediary step: Kernel-based classification}

\begin{figure}[H]
    \centering
    \begin{subfigure}[b]{0.48\linewidth}
\includegraphics[width=\linewidth]{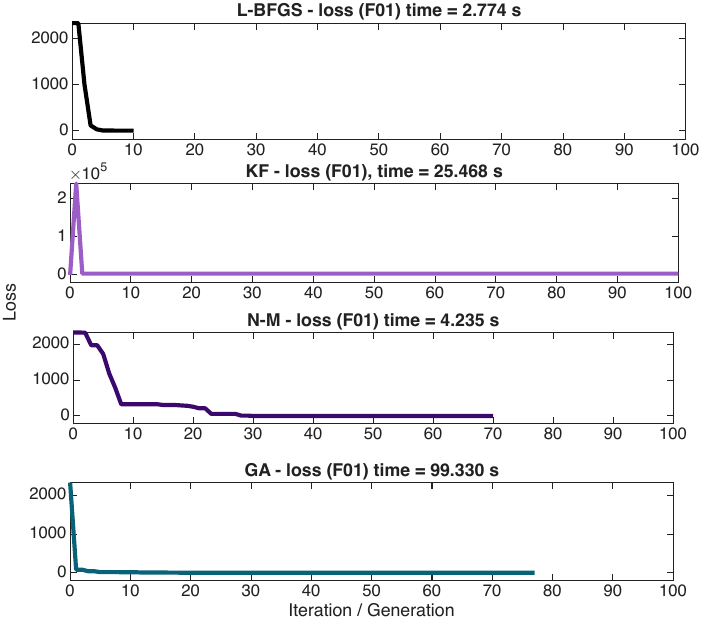}
    \caption{}
    \label{fig:lossGPR}
    \end{subfigure}
    \centering
    \begin{subfigure}[b]{0.48\linewidth}
\includegraphics[width=\linewidth]{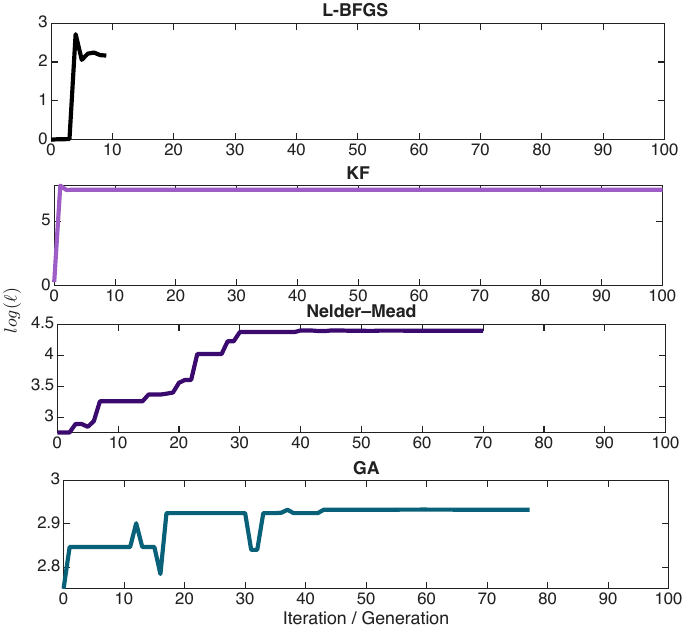}
    \caption{}
    \label{fig:GPRconvergenceParams}
    \end{subfigure}
    \caption{Deterministic kernel parameter optimization (a) loss trace and (b) kernel parameter convergence for fault \textbf{F01}. All four converged in a perfect fault detection in the control charts.}
    \label{fig:GPRcalibration}
\end{figure}

Figure \ref{fig:GPRcalibration} summarizes deterministic parameter optimisation for the GPC model with an SE kernel (fault \textbf{F01}). All four optimisers converged to solutions yielding perfect fault detection in the subsequent control charts. Among the tested methods, L-BFGS was the most computationally efficient, requiring 2.775 s until automatic convergence, whereas the GA was substantially slower (99.330 s).

The convergence behaviour differs across optimisers (Fig. \ref{fig:lossGPR}–\ref{fig:GPRconvergenceParams}). GA and Nelder–Mead converged to similar kernel lengthscale values. Kernel Flows exhibits a smooth initial trajectory, followed by noticeable oscillations near convergence. Nelder–Mead shows a characteristic stepwise evolution, with the lengthscale progressing through several plateaus before settling to the final value.

Although all optimisation methods achieved strong convergence in the classification setting \ref{tab:gpr_opt_methods}, good classification performance does not necessarily translate into optimal behaviour in the subsequent control charts. This highlights the distinction between kernel calibration for supervised discrimination and calibration for statistical process monitoring.

In previous work \cite{DUMA2026103679}, Kernel Flows was shown to be an effective strategy for learning kernel parameters in the context of Kernel MSPC via K-PCR. However, for GPC, the results in Table~\ref{tab:gpr_opt_methods} indicate a slightly different ranking. Based on the mean performance across faults, the overall ordering is GA $>$ N-M $>$ L-BFGS $>$ KF. Nevertheless, the performance differences between GA, N–M, and L-BFGS are marginal.

Given the substantial computational cost of GA (approximately 60 seconds) compared to L-BFGS (approximately 0.6 seconds), the small improvement in average performance does not justify the additional computational burden in the GPC setting. Consequently, L-BFGS was selected as the preferred optimiser for the supervised calibration stage using GPC as a classifier. It is also worth noting that many fault results are identical across optimisation methods. Only a few cases (notably \textbf{F13}, \textbf{F15}, \textbf{F16}, and \textbf{F18}) show slight improvements for Nelder–Mead and GA compared to L-BFGS. Overall, the results suggest that first-order gradient-based optimisation provides an efficient and sufficiently accurate solution for kernel parameter estimation in the GPC-based intermediary step.

\begin{table}[H]
\centering
\renewcommand{\arraystretch}{1.15}
\setlength{\tabcolsep}{3pt}
\caption{GPC optimization with an SE Kernel. Performance comparison of four parameter optimisation methods (LBFGS, Kernel Flows, Nelder--Mead, Genetic Algorithm) across 21 TEP faults. Metrics include AUC, F1-score, False Alarm Rate (FAR), and Fault Detection Rate (FDR).}
\label{tab:gpr_opt_methods}
\begin{tabular}{c|cccc|cccc|cccc|cccc}
\hline
\multirow{2}{*}{Fault} &
\multicolumn{4}{c|}{\textbf{L-BFGS}} &
\multicolumn{4}{c|}{\textbf{Kernel Flows}} &
\multicolumn{4}{c|}{\textbf{Nelder-Mead}} &
\multicolumn{4}{c}{\textbf{Genetic Algorithm}} \\
& AUC & F1 & FAR & FDR & AUC & F1 & FAR & FDR & AUC & F1 & FAR & FDR & AUC & F1 & FAR & FDR \\
\hline
\textbf{F01} & 1.00 & 1.00 & 0.00 & 1.00 & 1.00 & 1.00 & 0.00 & 1.00 & 1.00 & 1.00 & 0.00 & 1.00 & 1.00 & 1.00 & 0.00 & 1.00 \\
\textbf{F02} & 0.99 & 0.98 & 0.00 & 0.96 & 0.99 & 0.98 & 0.00 & 0.95 & 0.99 & 0.98 & 0.00 & 0.97 & 0.99 & 0.98 & 0.00 & 0.97 \\
\textbf{F03} & 0.49 & 0.58 & 0.46 & 0.45 & 0.44 & 0.00 & 0.00 & 0.00 & 0.50 & 0.63 & 0.54 & 0.51 & 0.50 & 0.64 & 0.55 & 0.53 \\
\textbf{F04} & 1.00 & 1.00 & 0.01 & 1.00 & 0.50 & 0.00 & 0.00 & 0.00 & 1.00 & 1.00 & 0.00 & 1.00 & 1.00 & 1.00 & 0.00 & 1.00 \\
\textbf{F05} & 1.00 & 1.00 & 0.00 & 1.00 & 0.39 & 0.00 & 0.00 & 0.00 & 1.00 & 1.00 & 0.00 & 1.00 & 1.00 & 1.00 & 0.00 & 1.00 \\
\textbf{F06} & 1.00 & 1.00 & 0.00 & 1.00 & 1.00 & 0.91 & 0.00 & 0.83 & 1.00 & 1.00 & 0.00 & 1.00 & 1.00 & 1.00 & 0.00 & 1.00 \\
\textbf{F07} & 1.00 & 1.00 & 0.00 & 1.00 & 1.00 & 1.00 & 0.00 & 1.00 & 1.00 & 1.00 & 0.00 & 1.00 & 1.00 & 1.00 & 0.00 & 1.00 \\
\textbf{F08} & 0.99 & 0.86 & 0.00 & 0.75 & 0.91 & 0.26 & 0.00 & 0.15 & 0.99 & 0.88 & 0.00 & 0.78 & 0.99 & 0.88 & 0.00 & 0.78 \\
\textbf{F09} & 0.68 & 0.54 & 0.15 & 0.38 & 0.60 & 0.00 & 0.00 & 0.00 & 0.66 & 0.57 & 0.18 & 0.41 & 0.66 & 0.57 & 0.18 & 0.41 \\
\textbf{F10} & 0.81 & 0.73 & 0.18 & 0.60 & 0.24 & 0.00 & 0.00 & 0.00 & 0.84 & 0.80 & 0.19 & 0.70 & 0.84 & 0.80 & 0.19 & 0.70 \\
\textbf{F11} & 0.72 & 0.62 & 0.18 & 0.46 & 0.37 & 0.00 & 0.00 & 0.00 & 0.74 & 0.69 & 0.19 & 0.54 & 0.74 & 0.69 & 0.19 & 0.55 \\
\textbf{F12} & 0.98 & 0.88 & 0.00 & 0.79 & 0.98 & 0.88 & 0.00 & 0.78 & 0.99 & 0.88 & 0.00 & 0.78 & 0.99 & 0.87 & 0.00 & 0.78 \\
\textbf{F13} & 0.81 & 0.28 & 0.00 & 0.16 & 0.78 & 0.17 & 0.01 & 0.09 & 0.79 & 0.63 & 0.00 & 0.46 & 0.79 & 0.63 & 0.00 & 0.46 \\
\textbf{F14} & 0.99 & 0.97 & 0.03 & 0.94 & 0.48 & 0.00 & 0.00 & 0.00 & 1.00 & 0.97 & 0.03 & 0.96 & 1.00 & 0.97 & 0.03 & 0.96 \\
\textbf{F15} & 0.68 & 0.53 & 0.12 & 0.37 & 0.39 & 0.00 & 0.00 & 0.00 & 0.70 & 0.63 & 0.14 & 0.48 & 0.70 & 0.64 & 0.14 & 0.49 \\
\textbf{F16} & 0.78 & 0.77 & 0.22 & 0.65 & 0.40 & 0.00 & 0.00 & 0.00 & 0.82 & 0.83 & 0.28 & 0.75 & 0.82 & 0.83 & 0.28 & 0.75 \\
\textbf{F17} & 0.95 & 0.86 & 0.03 & 0.76 & 0.35 & 0.00 & 0.00 & 0.00 & 0.95 & 0.88 & 0.03 & 0.78 & 0.95 & 0.88 & 0.03 & 0.79 \\
\textbf{F18} & 0.38 & 0.39 & 0.00 & 0.24 & 0.24 & 0.03 & 0.03 & 0.01 & 0.51 & 0.52 & 0.00 & 0.35 & 0.52 & 0.52 & 0.00 & 0.35 \\
\textbf{F19} & 0.55 & 0.78 & 0.84 & 0.75 & 0.27 & 0.17 & 0.01 & 0.09 & 0.63 & 0.80 & 0.76 & 0.78 & 0.70 & 0.81 & 0.63 & 0.77 \\
\textbf{F20} & 0.88 & 0.75 & 0.02 & 0.60 & 0.24 & 0.00 & 0.00 & 0.00 & 0.88 & 0.76 & 0.03 & 0.61 & 0.88 & 0.76 & 0.03 & 0.62 \\
\textbf{F21} & 0.48 & 0.22 & 0.06 & 0.13 & 0.32 & 0.00 & 0.00 & 0.00 & 0.49 & 0.23 & 0.08 & 0.13 & 0.49 & 0.23 & 0.08 & 0.13 \\
\hline
\hline
\textbf{Mean} 
& 0.82 & 0.75 & 0.11 & 0.67
& 0.57 & 0.26 & 0.00 & 0.23
& 0.83 & 0.79 & 0.12 & 0.71
& 0.84 & 0.80 & 0.11 & 0.72 \\
\hline
\end{tabular}
\end{table}

\subsubsection{Deterministic Control Charts: Supervised Calibration}

For Fault~F01, although perfect classification performance was achieved in the supervised GPC stage, the corresponding control-chart behaviour reveals important differences. In particular, convergence of the kernel parameters for the classifier does not imply equivalent performance when the same parameters are used in a K-PCA--based MSPC framework.

\begin{figure}[H]
    \centering
    \begin{subfigure}[b]{0.49\linewidth}
        \centering
        \includegraphics[width=\linewidth]{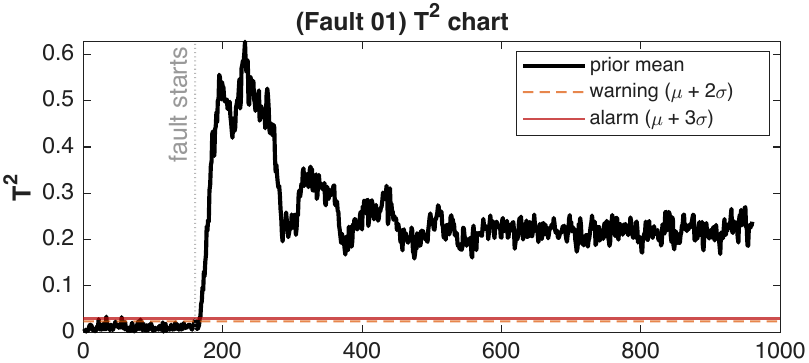}
        \caption{}
        \label{fig:nonpt2}
    \end{subfigure}
    \hfill
    \begin{subfigure}[b]{0.49\linewidth}
        \centering
        \includegraphics[width=\linewidth]{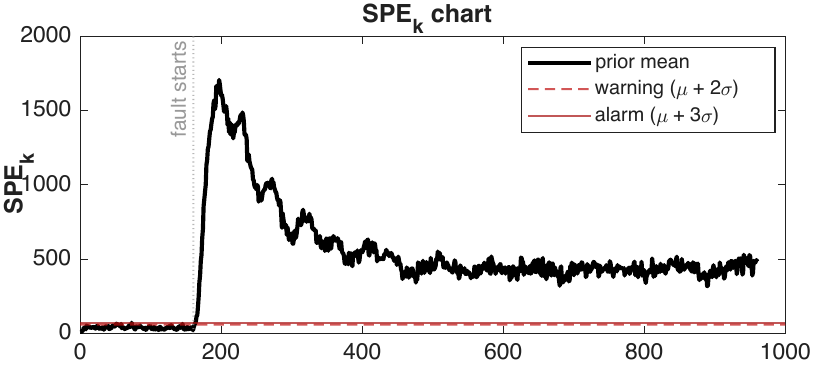}
        \caption{}
        \label{fig:nonpspex}
    \end{subfigure}
    \hfill
    \begin{subfigure}[b]{0.49\linewidth}
        \centering
        \includegraphics[width=\linewidth]{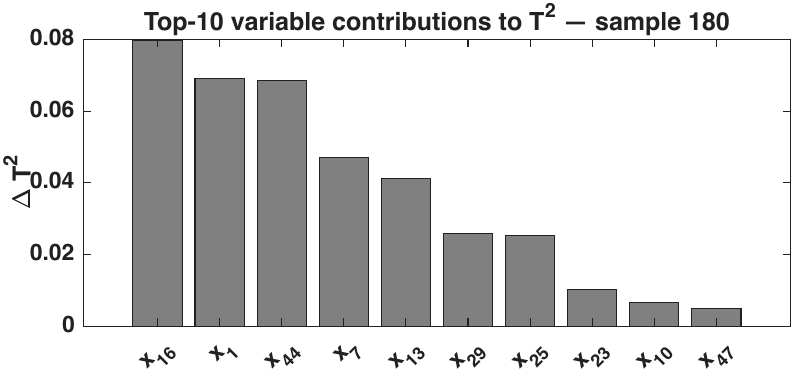}
        \caption{}
        \label{fig:nonptecontr}
    \end{subfigure}
    \hfill
    \begin{subfigure}[b]{0.49\linewidth}
        \centering
        \includegraphics[width=\linewidth]{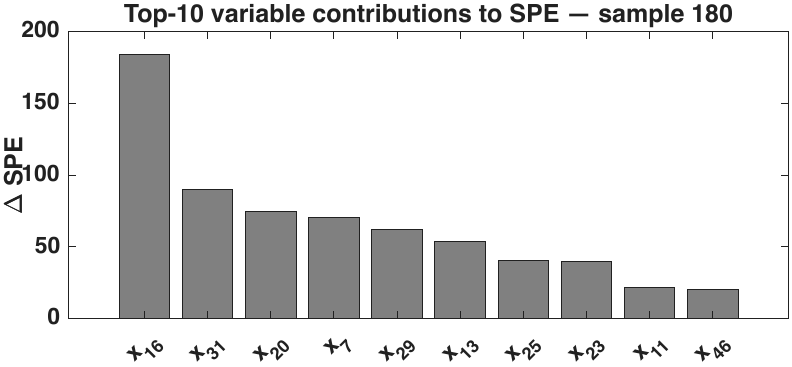}
        \caption{}
        \label{fig:nonpspexcontr}
    \end{subfigure}
    \caption{K-PCA MSPC control charts calibrated with the GPC-L-BFGS convergence kernel parameters, for a SE kernel.}
    \label{fig:deterministic}
\end{figure}

When applying K-PCA MSPC with kernel parameters optimised via the GPC--L-BFGS route, the resulting deterministic monitoring statistics (Figure~\ref{fig:deterministic}a-b) detect the onset of the fault correctly. The contribution analysis for a sample taken 80 minutes after the fault strart (Figure~\ref{fig:deterministic}c-d) provides further insight. Variables $\mathbf{x}_{16}$, $\mathbf{x}_{1}$ and $\mathbf{x}_{44}$ are the primary drivers of the out-of-control behaviour in the $T^2$ statistic, indicating that the dominant deviations lie within the principal subspace captured by the K-PCA model. Variable $\mathbf{x}_{16}$ presents also new variation, as seen in the contribution chart for explaining the $SPE_k$ value at timestamp 180. 

\subsubsection{MCMC Results}

\begin{figure}[H]
    \centering
    \begin{subfigure}[b]{0.99\linewidth}
        \includegraphics[width=\linewidth]{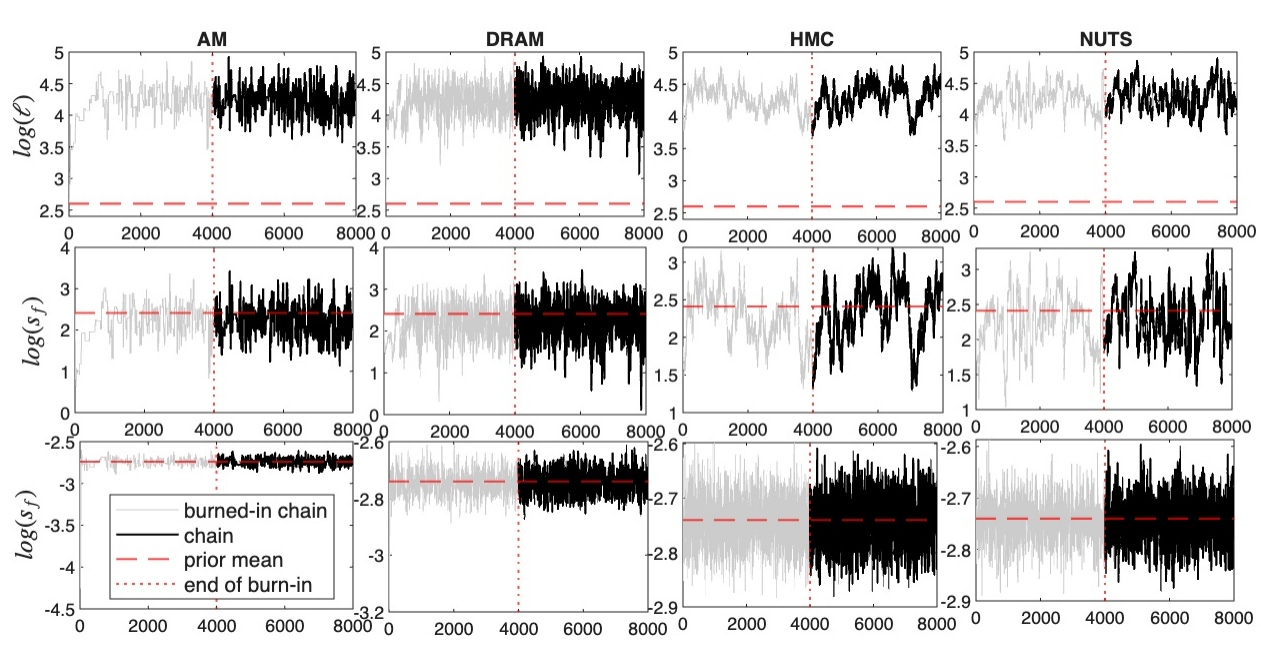}
        \caption{}
        \label{fig:chins}
    \end{subfigure}
    \begin{subfigure}[b]{0.9\linewidth}
        \includegraphics[width=\linewidth]{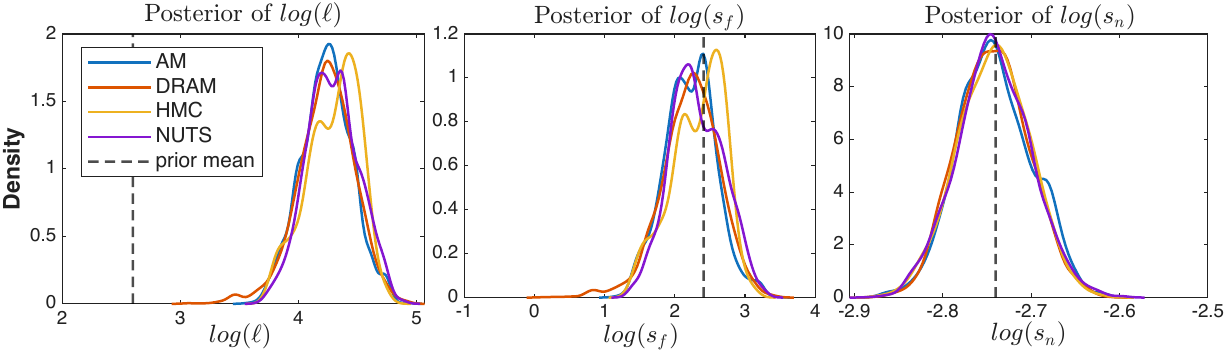}
    \end{subfigure}
    \caption{MCMC (a) chains for the GPC-based log-likelihood, with L-BFGS-GPC convergence values as prior mean and (b) the posterior distributions. }
    \label{fig:chains}
\end{figure}

Figure~\ref{fig:chains} presents the MCMC chains obtained for the SE kernel using four sampling strategies: AM, DRAM, HMC, and the NUTS. The prior means were set to the deterministic L-BFGS estimates from the GPC optimisation stage.

In terms of convergence behaviour, AM and DRAM exhibit similar posterior trajectories and stabilise around comparable parameter regions. Likewise, HMC and NUTS converge to closely aligned posterior distributions, reflecting their shared gradient-based sampling structure. While all samplers reach stationarity, the gradient-based methods demonstrate smoother exploration of the parameter space. For the kernel lengthscale parameter $\log(\ell)$, all four samplers converge to posterior modes that are larger than the prior mean, and in a similar MAP value. This indicates that the probabilistic calibration favours a smoother kernel (i.e., larger lengthscale) than suggested by the deterministic optimisation alone. For the signal variance parameter $\log(s_f)$, a clear distinction is observed between sampling families. NUTS and DRAM converge to posterior values that are slightly smaller than the prior mean, suggesting a reduction in effective signal amplitude when uncertainty is fully accounted for. In contrast, AM yields posterior estimates that remain close to the prior mean, indicating weaker adjustment of this parameter. Finally, for the noise variance parameter $\log(s_n)$, all four sampling approaches converge near the prior mean. This suggests that the deterministic estimate already provides a stable and well-identified value for the noise level, with limited posterior correction required.

ESS was computed for each parameter using the post-burn-in chains. The results indicate that DRAM achieves the highest sampling efficiency, with ESS values between approximately 305 and 486 across parameters. The AM sampler produces moderately efficient chains, with ESS values around 188–223. In contrast, the gradient-based samplers show more uneven efficiency. For HMC, the ESS for the kernel parameters $\log(\ell)$ and $\log(s_f)$ is relatively low (approximately 18), indicating strong autocorrelation in these chains, although the noise variance parameter $\log(s_n)$ mixes efficiently (ESS = 833). NUTS improves upon HMC but still yields lower ESS values for $\log(\ell)$ and $\log(s_f)$ (= 41) compared to the adaptive random-walk samplers.

Considering the minimum ESS across parameters, DRAM provides the most efficient exploration of the posterior distribution, followed by AM, while HMC and NUTS exhibit slower mixing for the kernel parameters. These results suggest that, for this posterior geometry, adaptive Metropolis-type samplers provide more robust sampling performance than the gradient-based methods under the current tuning settings.

The MCMC results for Kernel PCR are presented in \textbf{\ref{app:KPCRResults}}. In contrast to the GPC-based optimisation, the AM and DRAM samplers no longer converge to the same posterior distribution of the kernel parameters. Instead, DRAM, HMC, and NUTS converge to posterior distributions centred around a similar value of $\log(\ell)$, whereas AM explores a slightly different region of the parameter space.

In terms of convergence speed, HMC exhibits the fastest stabilisation, reaching a stationary regime after approximately 100 iterations. NUTS converges shortly thereafter, around 250 iterations, while DRAM requires a longer burn-in period of roughly 1000 iterations before stabilising. This highlights the improved efficiency of gradient-based samplers for the K-PCR likelihood landscape.

\subsubsection{Probabilistic MSPC control charts: supervised calibration}

\begin{figure}[H]
    \centering
    
    \begin{subfigure}[b]{0.65\linewidth}      \includegraphics[width=\linewidth]{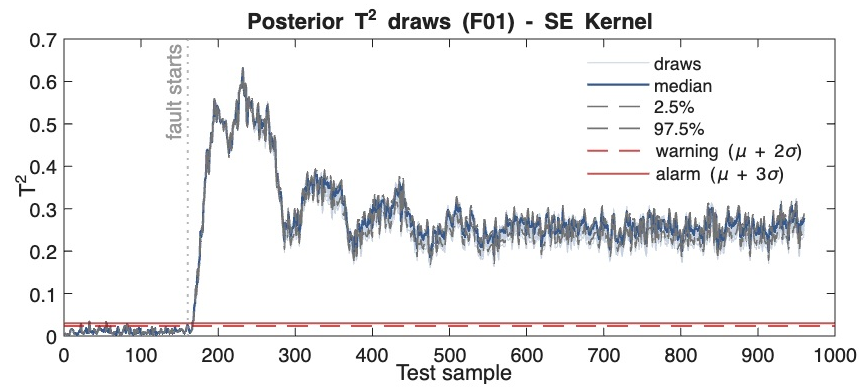}
        \caption{}
        \label{fig:T2SE}
    \end{subfigure}

    \begin{subfigure}[b]{0.65\linewidth}   \includegraphics[width=\linewidth]{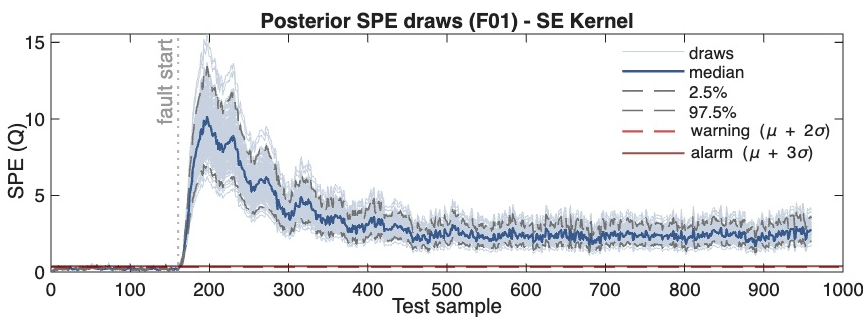}
        \caption{}
        \label{fig:SPExSE}
    \end{subfigure}
    
    \begin{subfigure}[b]{0.73\linewidth}  
    \centering
    \includegraphics[width=\linewidth]{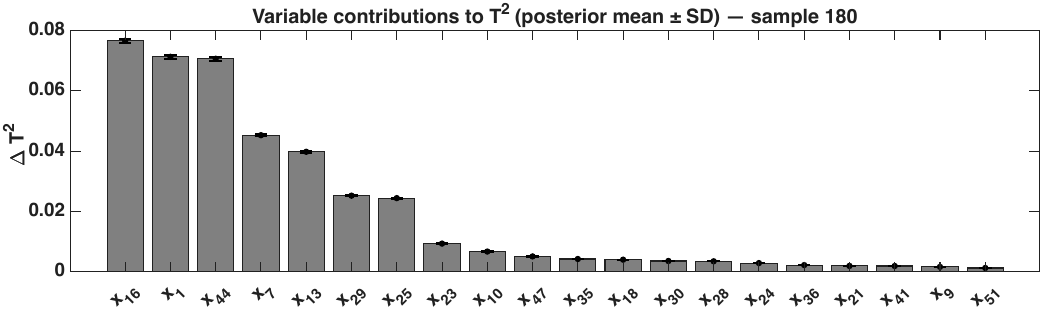}
        \caption{}
        \label{fig:t2SEcontr}
    \end{subfigure}
    \begin{subfigure}[b]{0.7\linewidth}   
    \centering
    \includegraphics[width=\linewidth]{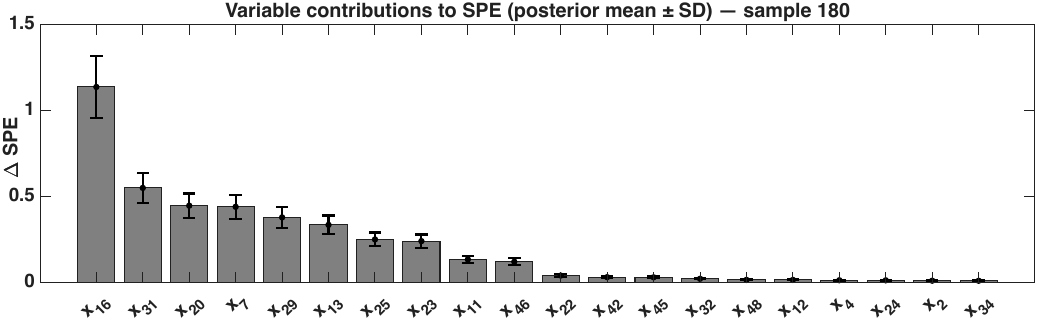}
        \caption{}
        \label{fig:SPExSEcontr}
    \end{subfigure}
    \caption{Probabilistic control charts (a,b) and variable contributions to the control charts (c, d) for the SE Kernel K-PCA, with DRAM MCMC sampling and GPC-based prior mean.}
    \label{fig:probabilisticControlCharts}
\end{figure}

\begin{figure}[H]
    \centering
    \begin{subfigure}[b]{0.45\linewidth}
\includegraphics[width=\linewidth]{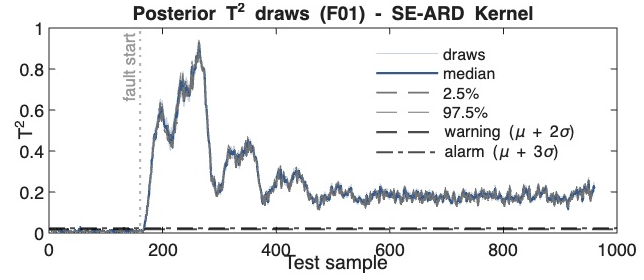}
    \caption{}
    \label{fig:ARDSET2chart}
    \end{subfigure}
    \begin{subfigure}[b]{0.45\linewidth}
\includegraphics[width=\linewidth]{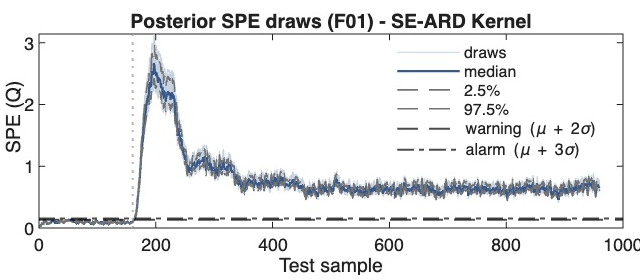}
    \caption{}
    \label{fig:ARDSESPExChart}
    \end{subfigure}
    \begin{subfigure}[b]{0.45\linewidth}
\includegraphics[width=\linewidth]{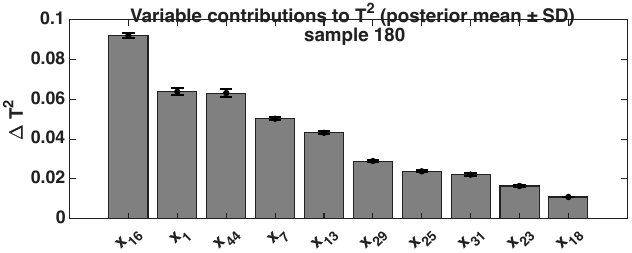}
    \caption{}
    \label{fig:ARDSET2Contr}
    \end{subfigure}
    \begin{subfigure}[b]{0.45\linewidth}
\includegraphics[width=\linewidth]{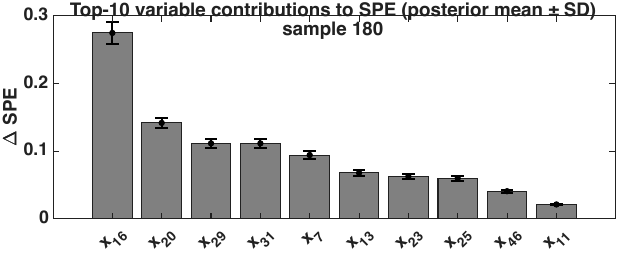}
    \caption{}
    \label{fig:ARDSESPExContr}
    \end{subfigure}
    \caption{Probabilistic control charts (a,b) and variable contributions to the control charts (c, d) for the SE-ARD kernel, with DRAM sampling and GPC-based prior mean. Similar visualisations for K-PCR are found in \textbf{\ref{app:KPCRResults}}.}
\label{fig:probabisticArdSeControlCharts}
\end{figure}

Figures~\ref{fig:probabilisticControlCharts} and 
\ref{fig:probabisticArdSeControlCharts} present the probabilistic $T^2$ and SPE control charts, together with their corresponding contribution plots, for the SE and ARD-SE kernels, respectively. A clear pattern is observed in the uncertainty bands. When the process operates within control limits (healthy region), the posterior uncertainty in both $T^2$ and SPE is very small, indicating high confidence in the in-control state. However, once the fault is initiated, the uncertainty increases substantially. This reflects the epistemic uncertainty in kernel parameters being amplified under abnormal process conditions.

Moreover, the uncertainty is consistently larger for the SPE control chart than for the $T^2$ chart. This behaviour is expected, as SPE captures residual variation outside the modeled K-PCA subspace and is therefore more sensitive to model misspecification and parameter variability. In contrast, $T^2$ represents variation within the retained principal component subspace, which is typically more stable. The same behaviour is observed in the contribution plots. Variable contributions to the SPE chart exhibit noticeably wider uncertainty bands than those associated with the $T^2$ chart. This indicates that uncertainty in kernel parameters propagates more strongly to residual-space diagnostics than to subspace-based variation.

A relationship between contribution magnitude and uncertainty can be observed: variables with lower contribution values tend to exhibit smaller posterior uncertainty, whereas variables driving the out-of-control behaviour display both larger contributions and wider uncertainty intervals. This suggests that uncertainty propagation is proportional to the strength of a variable's influence on the monitoring statistic. Overall, the probabilistic formulation not only improves fault detection performance but also provides additional interpretability by quantifying confidence in both detection and variable attribution.

In \ref{app:detvsprob}, Table \ref{tab:SEdetvsprob} compares the fault detection performance of the deterministic control charts (prior mean of the kernel parameters) with the posterior-mean-based charts for the SE kernel. The comparison is made in terms of the composite indicator (CI), which summarises the trade-off between false alarm rate (FAR) and fault detection rate (FDR).

The results show that nearly all faults benefit from the probabilistic treatment of the kernel parameters. In several cases, the improvement is substantial. For instance, Fault \textbf{F01} improves from $\mathrm{CI}=0.82$ to $0.99$, \textbf{F04} from $0.57$ to $0.75$, \textbf{F14} from $0.53$ to $0.92$, \textbf{F17} from $0.60$ to $0.89$, \textbf{F20} from $0.61$ to $0.71$, and \textbf{F21} from $0.50$ to $0.81$. These results indicate that incorporating posterior uncertainty into the control chart construction can significantly enhance fault detectability, particularly for faults that are only partially captured under the deterministic (prior-mean) setting.

When employing the ARD-SE kernel (Figures~\ref{fig:probabisticArdSeControlCharts}), several noteworthy differences can be observed in comparison with the isotropic SE kernel. The overall uncertainty in both the $T^2$ and SPE control charts is reduced. The posterior bands become narrower, indicating that the additional flexibility introduced by ARD allows the model to better adapt to variable-specific scaling. Nevertheless, the general uncertainty pattern remains unchanged: uncertainty is small during normal operation and increases after the onset of the fault. This confirms that the larger posterior variability is primarily associated with abnormal operating conditions rather than an artifact of kernel choice.

For the $T^2$ contribution plots, the ranking of the most influential variables remains largely unchanged. However, their relative magnitudes are modified. In particular, variable $\mathbf{x}_{16}$ becomes the dominant contributor to the out-of-control behaviour. This suggests that ARD redistributes variance within the modeled subspace by reweighting individual variable lengthscales, thereby refining the attribution without fundamentally altering the diagnostic structure. In the SPE contribution charts, the posterior uncertainty is again reduced compared to the isotropic SE case. At the same time, minor changes in the top-10 ranking of contributing variables can be observed. This indicates that residual-space diagnostics are more sensitive to anisotropic scaling, as ARD modifies how unexplained variation is allocated across variables.

Overall, the ARD-SE kernel preserves the qualitative diagnostic conclusions while reducing posterior uncertainty and slightly refining variable attribution. This highlights the benefit of variable-specific lengthscales in stabilizing probabilistic control charts without altering the underlying fault interpretation.

With respect to the ARD-SE kernel, the comparison between deterministic (prior mean) and posterior-mean control charts is presented in Table \ref{tab:ARDseCI} in \textbf{\ref{app:detvsprob}}. In contrast to the SE kernel case, only marginal differences are observed between the deterministic and probabilistic approaches. That is, incorporating posterior uncertainty in the ARD-SE setting does not lead to substantial additional improvements in the composite indicator.

Nevertheless, both the deterministic and probabilistic ARD-SE results exhibit a clear performance improvement compared to the corresponding SE kernel results for most faults. This suggests that the increased flexibility of the ARD-SE kernel already captures a significant portion of the model uncertainty through its additional lengthscale parameters. An exception is fault \textbf{F21}, for which the SE kernel configuration yields better performance than the ARD-SE setup.

\subsection{Unsupervised Calibration}

Table~\ref{tab:unsup-kernel-methods-composite} presents the composite indicator results for the unsupervised deterministic kernel calibration, where only the SE kernel parameters were optimised. With the exception of methods \textbf{M4}, \textbf{M5}, and \textbf{M7}, which consistently converge to values around 0.50 and therefore fail to distinguish between healthy and faulty operating conditions, the remaining approaches exhibit very similar performance levels. In particular, methods \textbf{M1}, \textbf{M2}, \textbf{M3}, \textbf{M6}, \textbf{M8}, \textbf{M9}, and \textbf{M10} achieve mean composite indicators in the range of 0.83–0.84, indicating stable and competitive fault detection performance across the 21 faults.

Interestingly, the average composite indicator obtained through several unsupervised strategies slightly exceeds the mean performance observed for both supervised SE and ARD-SE configurations with GPC (\textbf{\ref{app:detvsprob}}). This suggests that carefully designed unsupervised kernel selection criteria can provide highly competitive monitoring performance, even without access to labelled fault data.

Nevertheless, this global trend does not hold uniformly across all faults. Certain faults, such as \textbf{F21}, demonstrate improved performance under supervised calibration, indicating that the incorporation of labelled information can still be advantageous for specific fault structures.

\begin{table}[H]
\centering
\small
\caption{Composite score $\big((1-\mathrm{FAR})+\mathrm{FDR}\big)/2$ for 10 unsupervised kernel-tuning methods across 21 TEP faults (two decimals). The bottom row reports the mean score across faults for each method.}
\label{tab:unsup-kernel-methods-composite}
\setlength{\tabcolsep}{4pt}
\renewcommand{\arraystretch}{1.15}

\begin{tabular}{lcccccccccc}
\hline
\textbf{Fault} &
\textbf{M1} &
\textbf{M2} &
\textbf{M3} &
\textbf{M4} &
\textbf{M5} &
\textbf{M6} &
\textbf{M7} &
\textbf{M8} &
\textbf{M9} &
\textbf{M10} \\
\hline
\textbf{F01} & 0.99 & 0.99 & 0.99 & 0.50 & 0.50 & 0.99 & 0.49 & 0.99 & 0.99 & 0.99 \\
\textbf{F02} & 0.98 & 0.98 & 0.98 & 0.50 & 0.50 & 0.98 & 0.49 & 0.98 & 0.98 & 0.98 \\
\textbf{F03} & 0.53 & 0.53 & 0.53 & 0.50 & 0.50 & 0.53 & 0.49 & 0.51 & 0.51 & 0.53 \\
\textbf{F04} & 0.97 & 0.97 & 0.94 & 0.50 & 0.50 & 0.92 & 0.49 & 0.99 & 0.99 & 0.94 \\
\textbf{F05} & 0.66 & 0.65 & 0.65 & 0.50 & 0.50 & 0.63 & 0.49 & 0.64 & 0.64 & 0.64 \\
\textbf{F06} & 0.99 & 0.99 & 0.99 & 0.50 & 0.50 & 0.99 & 0.49 & 0.99 & 0.99 & 0.99 \\
\textbf{F07} & 0.99 & 0.99 & 0.99 & 0.50 & 0.50 & 0.99 & 0.49 & 0.99 & 0.99 & 0.99 \\
\textbf{F08} & 0.98 & 0.98 & 0.98 & 0.50 & 0.50 & 0.98 & 0.49 & 0.98 & 0.98 & 0.98 \\
\textbf{F09} & 0.49 & 0.50 & 0.49 & 0.50 & 0.50 & 0.49 & 0.49 & 0.51 & 0.51 & 0.49 \\
\textbf{F10} & 0.80 & 0.80 & 0.79 & 0.50 & 0.50 & 0.79 & 0.49 & 0.81 & 0.81 & 0.79 \\
\textbf{F11} & 0.84 & 0.86 & 0.82 & 0.50 & 0.50 & 0.85 & 0.49 & 0.87 & 0.87 & 0.83 \\
\textbf{F12} & 0.98 & 0.98 & 0.98 & 0.50 & 0.50 & 0.98 & 0.49 & 0.98 & 0.98 & 0.98 \\
\textbf{F13} & 0.97 & 0.97 & 0.96 & 0.50 & 0.50 & 0.97 & 0.49 & 0.97 & 0.97 & 0.97 \\
\textbf{F14} & 0.99 & 0.99 & 0.99 & 0.50 & 0.50 & 0.99 & 0.49 & 0.99 & 0.99 & 0.99 \\
\textbf{F15} & 0.63 & 0.62 & 0.61 & 0.50 & 0.50 & 0.62 & 0.50 & 0.60 & 0.60 & 0.60 \\
\textbf{F16} & 0.62 & 0.62 & 0.62 & 0.50 & 0.50 & 0.61 & 0.49 & 0.66 & 0.66 & 0.61 \\
\textbf{F17} & 0.96 & 0.96 & 0.95 & 0.50 & 0.50 & 0.95 & 0.49 & 0.97 & 0.97 & 0.95 \\
\textbf{F18} & 0.94 & 0.94 & 0.94 & 0.50 & 0.50 & 0.94 & 0.49 & 0.94 & 0.94 & 0.94 \\
\textbf{F19} & 0.59 & 0.58 & 0.59 & 0.50 & 0.50 & 0.60 & 0.49 & 0.71 & 0.71 & 0.61 \\
\textbf{F20} & 0.85 & 0.85 & 0.84 & 0.50 & 0.50 & 0.84 & 0.49 & 0.87 & 0.87 & 0.84 \\
\textbf{F21} & 0.70 & 0.71 & 0.71 & 0.50 & 0.50 & 0.70 & 0.49 & 0.72 & 0.72 & 0.70 \\
\hline
\textbf{Mean} 
& \textbf{0.83} & \textbf{0.83} & \textbf{0.83} & \textbf{0.50} & \textbf{0.50} 
& \textbf{0.83} & \textbf{0.49} & \textbf{0.84} & \textbf{0.84} & \textbf{0.83} \\
\hline
\end{tabular}
\end{table}

\begin{figure}[H]
    \centering
    \begin{subfigure}[b]{0.5\linewidth}
        \includegraphics[width=\linewidth]{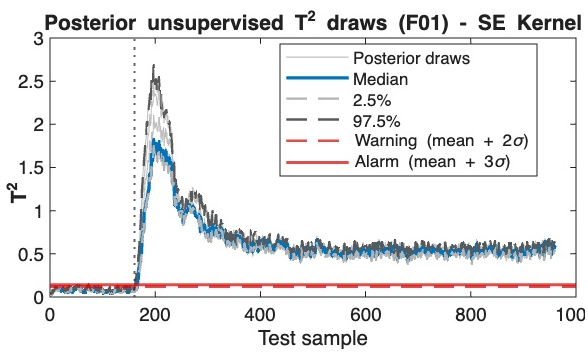}
        \caption{}
        \label{fig:unsupeT2}
    \end{subfigure}
    
    \begin{subfigure}[b]{0.5\linewidth}
        \includegraphics[width=\linewidth]{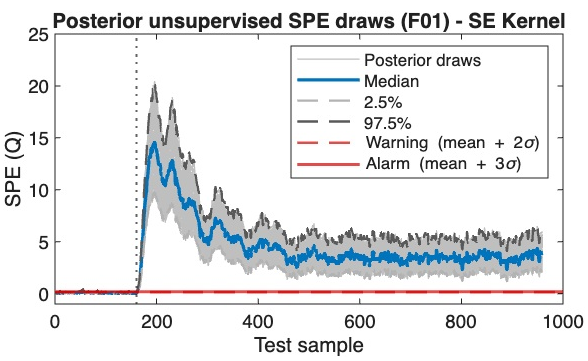}
        \caption{}
        \label{fig:unsupeT22}
    \end{subfigure}
    \caption{Probabilistic control charts for (a) T2 and (b) SPE metrics, using unsupervised kernel calibration (method M8).}
    \label{fig:unsupervisedControlCharts}
\end{figure}

The prior mean used for Bayesian sampling was obtained through the unsupervised \textbf{M8} method (MMD halves), which selects the kernel lengthscale by minimising the discrepancy between temporal partitions of healthy operation data. As illustrated in Figure~\ref{fig:unsupervisedControlCharts}, the uncertainty bands obtained under the unsupervised calibration are generally wider than those observed in the supervised setting. This reflects the additional epistemic uncertainty introduced when the kernel parameters are inferred without labelled fault information. 

Nevertheless, despite the increased posterior spread, the fault remains consistently detected across the entire distribution of posterior draws. In other words, the alarm condition is robust with respect to kernel parameter uncertainty, indicating stable fault discrimination even under unsupervised calibration.

\section{Summary and future work}

This work proposed a probabilistic extension of K-MSPC, with the goal of explicitly quantifying and propagating uncertainty in kernel parameters to the monitoring statistics. The framework follows a two-stage workflow: (i) deterministic kernel calibration to obtain informative prior means and (ii) Bayesian inference of kernel parameters using MCMC, yielding probabilistic control charts for both $T^2$ and SPE and their corresponding contribution plots.

In the supervised calibration setting, deterministic optimisation of an SE-kernel GPC model produced high classification performance across several optimisers. However, the study highlighted that strong convergence in a supervised classification objective does not necessarily translate into optimal kernel settings for K-PCA monitoring. Nevertheless, due to its favourable accuracy--runtime trade-off, L-BFGS was selected as the preferred deterministic optimiser for the supervised GPC intermediary step.

In the probabilistic stage, uncertainty in kernel parameterswas shown to propagate non-trivially to the control charts. For the SE kernel, posterior-mean control charts generally improved fault detection compared to deterministic prior-mean charts, with notable increases in the composite indicator for several faults (e.g., \textbf{F01}, \textbf{F14}, \textbf{F17}, and \textbf{F21}). The uncertainty bands were consistently wider during faulty operating regimes and narrow in the in-control region, indicating that parameter uncertainty is amplified under abnormal process behaviour. The residual-based SPE chart exhibited larger uncertainty than $T^2$, and the same pattern was reflected in the contribution plots, where variables with higher contributions also displayed wider posterior uncertainty.

When employing an ARD-SE kernel, the uncertainty in both $T^2$ and SPE charts decreased compared to the isotropic SE case, while preserving the qualitative uncertainty structure (small uncertainty in-control and larger uncertainty during faults). Variable-attribution patterns were largely preserved for $T^2$, while modest changes in contribution magnitudes and top-ranking variables were observed, particularly in the SPE contributions. In contrast to the SE kernel case, the deterministic and probabilistic ARD-SE results were very similar, suggesting that the additional flexibility of ARD already absorbs a portion of the modelling uncertainty in the deterministic stage.

Finally, an unsupervised calibration path was evaluated using a suite of deterministic kernel selection criteria. Several unsupervised methods achieved competitive average composite indicators across the benchmark faults. When the \textbf{M8} strategy (MMD halves) was used to define the prior mean in the probabilistic workflow, the resulting posterior control charts exhibited wider uncertainty than in the supervised setting, but still achieved robust fault detection across the full set of posterior draws. This demonstrates that probabilistic monitoring is feasible even in the absence of labelled fault data, while providing an explicit quantification of epistemic uncertainty.

Several directions are recommended to extend the proposed probabilistic K-MSPC framework:

\begin{itemize}
    \item \textbf{A framework to include dynamic process monitoring capabilities.} This would extend the approach to incorporate process dynamics, for example by probabilistic kernel DPCA or dynamic K-PCA formulations, and evaluate whether uncertainty propagation changes under temporal modelling.
    
    \item \textbf{Faster posterior approximations.} A possibility is to replace MCMC with scalable Bayesian approximations such as variational inference or Laplace-type approximations, with the aim of enabling near real-time uncertainty-aware monitoring.
    
    \item \textbf{Broader kernel families and structure.} A gap in the present research is the investigation of alternative kernels beyond SE/ARD-SE (e.g., Mat\'ern, rational quadratic, polynomial, and additive or composite kernels) and quantify how kernel choice affects posterior uncertainty and fault interpretability.
    
    \item \textbf{Calibration objectives aligned with monitoring.} A possibility is to develop deterministic kernel calibration criteria that directly target monitoring performance (e.g., control-limit stability, fault persistence capture, or separation in monitoring statistics), bridging the gap observed between classification-optimal and monitoring-optimal kernel settings.
    
    \item \textbf{Industrial case studies.} The TEP being a simulated dataset, the framework can be applied to real-world industrial datasets (e.g., spectroscopy and imaging-based monitoring), where sensor drift, missing data, and nonstationarity are present, and evaluate whether probabilistic charts improve robustness and trust in alarm decisions.
\end{itemize}

\section{Conclusions}\label{sc:conc}

This paper introduced a probabilistic extension of kernel-based multivariate statistical process control, aimed at making kernel parameter uncertainty explicit and actionable in nonlinear process monitoring. Instead of treating kernel parameters as fixed point estimates, the proposed framework infers their posterior distributions and propagates this epistemic uncertainty into the monitoring statistics.

A two-stage workflow was developed. Deterministic calibration (supervised via Gaussian process classification and K-PCR, or unsupervised via healthy-data criteria) provides informative prior means, after which Markov chain Monte Carlo sampling (AM, DRAM, HMC, NUTS) yields posterior draws of kernel parameters. These draws are propagated through kernel PCA to obtain probabilistic $T^2$ and SPE control charts and uncertainty-aware contribution plots.

Experiments on the Tennessee Eastman Process benchmark demonstrate that kernel parameter uncertainty can materially affect monitoring behaviour. For the isotropic SE kernel, posterior-mean charts frequently improve fault detection compared to deterministic prior-mean charts, and uncertainty bands are consistently small during healthy operation but widen after fault onset, indicating amplified epistemic uncertainty under abnormal regimes. SPE statistics exhibit larger posterior variability than $T^2$, and the same trend carries to contribution plots, where highly contributing variables also show wider credible intervals. For the ARD-SE kernel, posterior uncertainty is reduced and probabilistic and deterministic results are largely aligned, suggesting that the additional kernel flexibility absorbs part of the uncertainty already at the deterministic stage. Under unsupervised calibration, posterior uncertainty increases, yet fault detection remains robust across posterior draws.

Overall, the proposed probabilistic K-MSPC framework improves robustness and interpretability by attaching uncertainty to both fault detection and variable attribution, and it provides a principled route to uncertainty-aware monitoring in settings where kernel calibration is inherently uncertain.

\section*{Awknowledgements}
We acknowledge funding from the Flagship of Advanced Mathematics for Sensing, Imaging, and Modelling 2024--2031 (decision number 359183). VJ was supported through the Higher Education for Economic Transformation (HEET) program, funded by the \textbf{World Bank} through the Government of Tanzania.

\appendix

\section{Notation and deterministic K-PCA/K-MSPC background}\label{app:notation}

Let $\mathbf{X}\in\mathbb{R}^{n\times p}$ denote the autoscaled training data, let $\mathbf{K}$ denote the kernel matrix, and let $\mathbf{K}_c$ denote its centered version. The centered kernel matrix is eigendecomposed as
\[
\mathbf{K}_c\mathbf{U}=\mathbf{U}\boldsymbol{\Lambda},
\]
where $\boldsymbol{\Lambda}=\mathrm{diag}(\lambda_1,\dots,\lambda_r)$ contains the retained eigenvalues and $\mathbf{U}$ the corresponding eigenvectors.

For a new sample $\mathbf{x}$, the centered kernel vector is denoted by $\mathbf{k}_c(\mathbf{x})$, and the associated K-PCA score vector is
\[
\mathbf{t}(\mathbf{x})=\mathbf{k}_c(\mathbf{x})^\top \mathbf{U}\boldsymbol{\Lambda}^{-1/2}.
\]

The monitoring statistics are the Hotelling-type statistic
\[
T^2(\mathbf{x})=\sum_{h=1}^{r}\frac{t_h(\mathbf{x})^2}{\lambda_h},
\]
and the residual statistic
\[
\mathrm{SPE}(\mathbf{x})=k_c(\mathbf{x},\mathbf{x})-\sum_{h=1}^{r}t_h(\mathbf{x})^2.
\]

To support diagnosis, variable contribution diagnostics are also computed. The contribution of variable $d$ to the $T^2$ statistic is given by
\[
C_d^{T^2}(\mathbf{x})
=
\sum_{h=1}^{r}
\frac{t_h(\mathbf{x})}{\lambda_h}
\frac{\partial t_h(\mathbf{x})}{\partial x_d},
\]
which quantifies how variation in variable $d$ affects the modeled K-PCA subspace.

Similarly, the contribution of variable $d$ to the SPE statistic is defined as
\[
C_d^{\mathrm{SPE}}(\mathbf{x})
=
\frac{\partial \mathrm{SPE}(\mathbf{x})}{\partial x_d}.
\]

In the probabilistic formulation proposed in the present work, the same expressions are evaluated for each posterior draw of the kernel parameters, thereby yielding posterior distributions for both monitoring statistics and variable contributions.

\section{Deterministic vs. probabilistic results}\label{app:detvsprob}

\begin{table}[H]
\centering
\caption{False alarm rates (FAR) and fault detection rates (FDR) for T$^2$ and SPE charts (SE kernel), rounded to two decimals. Left: deterministic (prior mean). Right: posterior mean. CI is the composite indicator $\mathrm{CI}=\frac{(1-\mathrm{FAR})+\mathrm{FDR}}{2}$, averaged over T$^2$ and SPE.}
\label{tab:SEdetvsprob}
\begin{tabular}{c|ccccc|ccccc}
\hline
\textbf{Fault} 
& \multicolumn{5}{c|}{\textbf{Deterministic (prior mean)}} 
& \multicolumn{5}{c}{\textbf{Posterior mean}} \\
\textbf{} 
& \textbf{FAR$_{T^2}$} & \textbf{FDR$_{T^2}$} & \textbf{FAR$_{\text{SPE}}$} & \textbf{FDR$_{\text{SPE}}$} & \textbf{CI}
& \textbf{FAR$_{T^2}$} & \textbf{FDR$_{T^2}$} & \textbf{FAR$_{\text{SPE}}$} & \textbf{FDR$_{\text{SPE}}$} & \textbf{CI} \\
\hline
F01 & 0.03 & 0.66 & 0.01 & 0.66 & 0.82 & 0.01 & 0.99 & 0.01 & 1.00 &\textbf{ 0.99} \\
F02 & 0.02 & 0.98 & 0.01 & 0.98 & \textbf{0.98} & 0.00 & 0.96 & 0.01 & 0.99 &\textbf{ 0.98} \\
F03 & 0.98 & 0.99 & 0.97 & 1.00 & \textbf{0.51} & 0.01 & 0.00 & 0.00 & 0.00 & 0.50 \\
F04 & 0.03 & 0.16 & 0.02 & 0.16 & 0.57 & 0.02 & 0.15 & 0.00 & 0.88 &\textbf{ 0.75} \\
F05 & 0.01 & 0.25 & 0.01 & 0.25 & 0.62 & 0.02 & 0.28 & 0.00 & 0.26 & \textbf{0.63} \\
F06 & 0.06 & 0.97 & 0.02 & 0.97 & 0.97 & 0.02 & 0.99 & 0.00 & 1.00 & \textbf{0.99 }\\
F07 & 0.03 & 0.35 & 0.01 & 0.35 & 0.83 & 0.01 & 0.49 & 0.01 & 1.00 & \textbf{0.87} \\
F08 & 0.04 & 0.89 & 0.01 & 0.89 & 0.93 & 0.00 & 0.87 & 0.00 & 0.98 & \textbf{0.96} \\
F09 & 0.49 & 0.42 & 0.44 & 0.31 & 0.45 & 0.00 & 0.00 & 0.00 & 0.00 &\textbf{ 0.50} \\
F10 & 0.07 & 0.40 & 0.02 & 0.51 & \textbf{0.71} & 0.01 & 0.26 & 0.01 & 0.57 & 0.70 \\
F11 & 0.11 & 0.36 & 0.06 & 0.34 & 0.63 & 0.01 & 0.09 & 0.00 & 0.60 & \textbf{0.67} \\
F12 & 0.04 & 0.90 & 0.02 & 0.90 & 0.94 & 0.01 & 0.87 & 0.00 & 0.99 & \textbf{0.96} \\
F13 & 0.01 & 0.89 & 0.00 & 0.89 & 0.95 & 0.00 & 0.92 & 0.01 & 0.96 & \textbf{0.97} \\
F14 & 0.03 & 0.09 & 0.01 & 0.09 & 0.53 & 0.01 & 0.69 & 0.00 & 1.00 & \textbf{0.92} \\
F15 & 1.00 & 1.00 & 0.95 & 1.00 & 0.51 & 0.00 & 0.03 & 0.00 & 0.10 & \textbf{0.53} \\
F16 & 0.44 & 0.57 & 0.45 & 0.55 & \textbf{0.56} & 0.00 & 0.04 & 0.00 & 0.12 & 0.54 \\
F17 & 0.05 & 0.04 & 0.03 & 0.43 & 0.60 & 0.00 & 0.66 & 0.00 & 0.90 & \textbf{0.89} \\
F18 & 0.05 & 0.86 & 0.02 & 0.86 & 0.91 & 0.00 & 0.87 & 0.00 & 0.90 & \textbf{0.94} \\
F19 & 0.11 & 0.19 & 0.06 & 0.18 & \textbf{0.55} & 0.00 & 0.01 & 0.00 & 0.08 & 0.52 \\
F20 & 0.07 & 0.05 & 0.01 & 0.46 & 0.61 & 0.00 & 0.24 & 0.00 & 0.59 & \textbf{0.71} \\
F21 & 0.99 & 1.00 & 0.97 & 0.97 & 0.50 & 0.00 & 0.22 & 0.00 & 0.40 & \textbf{0.81} \\
\textbf{Mean} 
& \textbf{0.22} & \textbf{0.57} & \textbf{0.19} & \textbf{0.56} & \textbf{0.71}
& \textbf{0.01} & \textbf{0.46} & \textbf{0.00} & \textbf{0.68} & \textbf{0.79} \\
\hline

\hline
\end{tabular}
\end{table}

\begin{table}[H]
\centering
\caption{False alarm rates (FAR) and fault detection rates (FDR) for T$^2$ and SPE charts using an ARD-SE kernel, rounded to two decimals. Left: deterministic (prior mean). Right: posterior mean. Composite indicator (CI) is computed as $\mathrm{CI}=\frac{\mathrm{CI}_{T^2}+\mathrm{CI}_{\text{SPE}}}{2}$ with $\mathrm{CI}_\bullet=\frac{(1-\mathrm{FAR}_\bullet)+\mathrm{FDR}_\bullet}{2}$.}
\label{tab:ARDseCI}
\begin{tabular}{c|ccccc|ccccc}
\hline
\textbf{Fault} 
& \multicolumn{5}{c|}{\textbf{Deterministic (prior mean)}} 
& \multicolumn{5}{c}{\textbf{Posterior mean}} \\
\textbf{} 
& \textbf{FAR$_{T^2}$} & \textbf{FDR$_{T^2}$} & \textbf{FAR$_{\text{SPE}}$} & \textbf{FDR$_{\text{SPE}}$} & \textbf{CI}
& \textbf{FAR$_{T^2}$} & \textbf{FDR$_{T^2}$} & \textbf{FAR$_{\text{SPE}}$} & \textbf{FDR$_{\text{SPE}}$} & \textbf{CI} \\
\hline
F01 & 0.02 & 0.99 & 0.00 & 1.00 & 0.99 & 0.01 & 0.99 & 0.00 & 1.00 & 0.99 \\
F02 & 0.02 & 0.98 & 0.00 & 0.99 & 0.99 & 0.01 & 0.98 & 0.00 & 0.99 & 0.99 \\
F03 & 0.01 & 0.01 & 0.00 & 0.01 & 0.50 & 0.01 & 0.01 & 0.00 & 0.01 & 0.50 \\
F04 & 0.02 & 0.16 & 0.01 & 0.90 & 0.76 & 0.02 & 0.16 & 0.01 & 0.90 & 0.76 \\
F05 & 0.02 & 0.25 & 0.00 & 0.27 & 0.63 & 0.02 & 0.24 & 0.00 & 0.30 & 0.63 \\
F06 & 0.01 & 0.99 & 0.00 & 1.00 & 1.00 & 0.00 & 0.99 & 0.00 & 1.00 & 1.00 \\
F07 & 0.01 & 0.57 & 0.01 & 1.00 & 0.89 & 0.01 & 0.60 & 0.00 & 1.00 & \textbf{0.90} \\
F08 & 0.01 & 0.96 & 0.00 & 0.98 & 0.98 & 0.01 & 0.95 & 0.00 & 0.98 & 0.98 \\
F09 & 0.00 & 0.00 & 0.01 & 0.01 & 0.50 & 0.00 & 0.00 & 0.01 & 0.01 & 0.50 \\
F10 & 0.01 & 0.50 & 0.01 & 0.47 & \textbf{0.74} & 0.02 & 0.48 & 0.00 & 0.44 & 0.73 \\
F11 & 0.01 & 0.31 & 0.00 & 0.61 &\textbf{ 0.73} & 0.01 & 0.31 & 0.00 & 0.59 & 0.72 \\
F12 & 0.01 & 0.96 & 0.00 & 0.99 & 0.98 & 0.01 & 0.96 & 0.00 & 0.99 & 0.98 \\
F13 & 0.01 & 0.94 & 0.01 & 0.96 & 0.97 & 0.01 & 0.94 & 0.01 & 0.96 & 0.97 \\
F14 & 0.01 & 0.82 & 0.01 & 1.00 & 0.95 & 0.00 & 0.83 & 0.01 & 1.00 & \textbf{0.96} \\
F15 & 0.01 & 0.20 & 0.00 & 0.03 & 0.55 & 0.01 & 0.21 & 0.00 & 0.03 & \textbf{0.56} \\
F16 & 0.00 & 0.10 & 0.00 & 0.13 & \textbf{0.56} & 0.00 & 0.10 & 0.00 & 0.12 & 0.55 \\
F17 & 0.01 & 0.77 & 0.00 & 0.90 & \textbf{0.92} & 0.01 & 0.77 & 0.00 & 0.90 & 0.91 \\
F18 & 0.01 & 0.89 & 0.01 & 0.90 & 0.94 & 0.01 & 0.89 & 0.01 & 0.90 & 0.94 \\
F19 & 0.00 & 0.03 & 0.00 & 0.12 & 0.54 & 0.00 & 0.03 & 0.00 & 0.11 & 0.54 \\
F20 & 0.01 & 0.50 & 0.00 & 0.52 & 0.75 & 0.01 & 0.50 & 0.00 & 0.52 & 0.75 \\
F21 & 0.01 & 0.30 & 0.00 & 0.40 & 0.67 & 0.03 & 0.31 & 0.00 & 0.39 & 0.67 \\
\textbf{Mean} 
& \textbf{0.01} & \textbf{0.58} & \textbf{0.00} & \textbf{0.62} & \textbf{0.79}
& \textbf{0.01} & \textbf{0.58} & \textbf{0.00} & \textbf{0.62} & \textbf{0.79} \\

\hline
\end{tabular}
\end{table}

\section{K-PCR visualisations}\label{app:KPCRResults} 

\subsection{MCMC results for K-PCR}

\begin{figure}[H]
    \centering
    \begin{subfigure}[b]{0.99\linewidth}
\includegraphics[width=\linewidth]{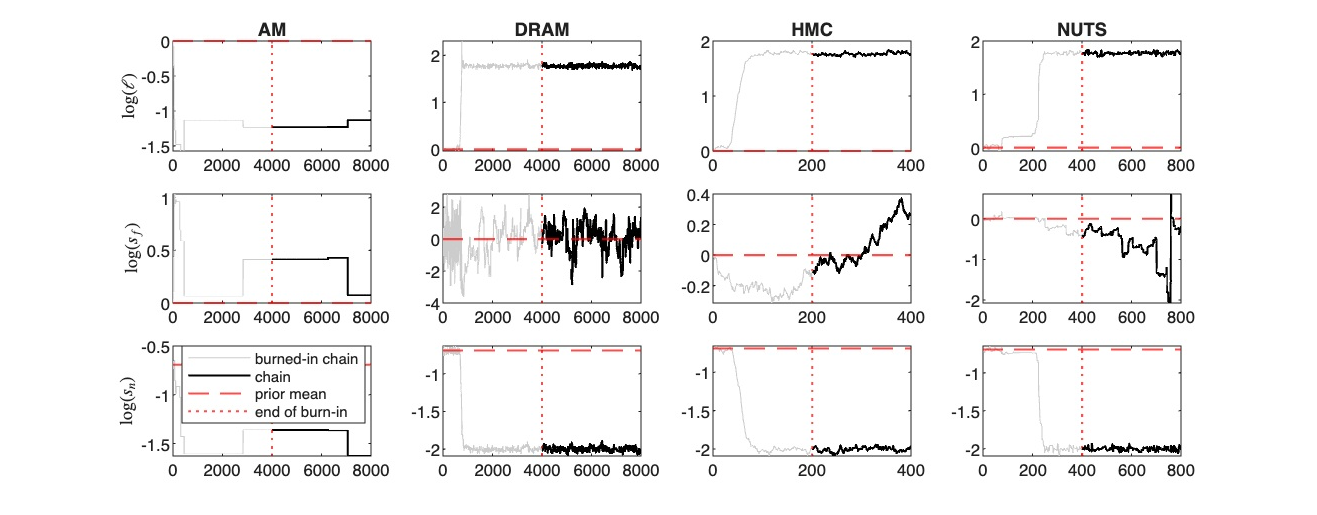}
        \caption{}
        \label{fig:B1}
    \end{subfigure}
    \begin{subfigure}[b]{0.9\linewidth}
\includegraphics[width=\linewidth]{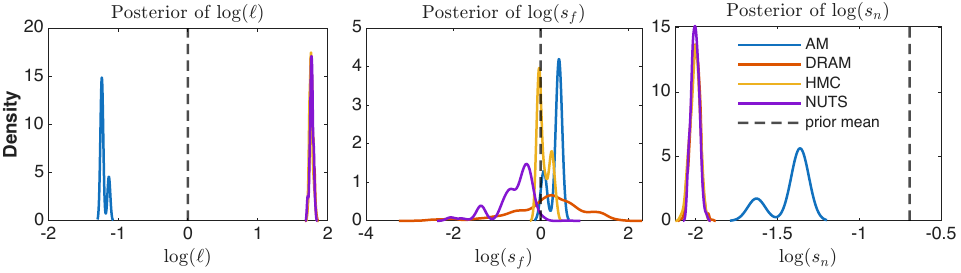}
        \caption{}
        \label{fig:B2}
    \end{subfigure}
    \caption{MCMC (a) chains for the K-PCR-based log-likelihood, with N-M convergence values as prior mean and (b) the posterior distribution of the SE kernel parameters.}
    \label{fig:K-PCRMCMC}
\end{figure}

\subsection{Probabilistic control charts for K-PCR}

\begin{figure}[H]
    \centering
    \begin{subfigure}[b]{0.6\linewidth}
        \includegraphics[width=\linewidth]{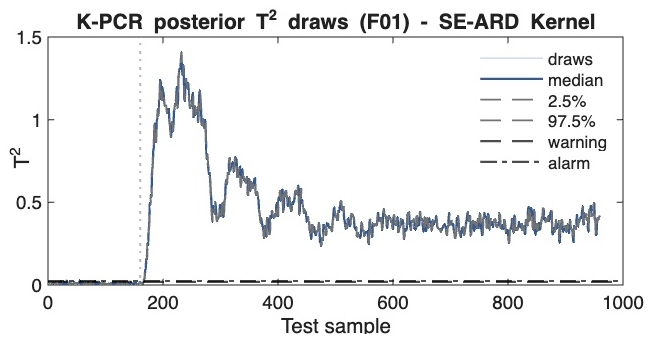}
        \caption{}
        \label{fig:B3T2}
    \end{subfigure}
    \begin{subfigure}[b]{0.6\linewidth}
        \includegraphics[width=\linewidth]{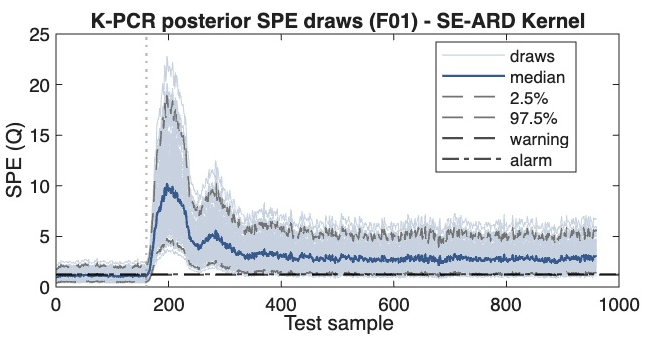}
        \caption{}
        \label{fig:B3SPE}
    \end{subfigure}
    \caption{The probabilistic control charts for the \textbf{F01} fault, using the SE-ARD kernel in a K-PCR prior mean estimation set-up.}
    \label{fig:K-PCR-controlCharts}
\end{figure}


\begin{thebibliography}{10}

\bibitem{Alam2014HyperparameterSI}
Md.~Ashad Alam and Kenji Fukumizu.
\newblock Hyperparameter selection in kernel principal component analysis.
\newblock {\em J. Comput. Sci.}, 10:1139--1150, 2014.

\bibitem{bach2002kernel}
Francis~R. Bach and Michael~I. Jordan.
\newblock Kernel independent component analysis.
\newblock {\em Journal of Machine Learning Research}, 3:1--48, 2002.

\bibitem{belkin2003laplacian}
Mikhail Belkin and Partha Niyogi.
\newblock Laplacian eigenmaps for dimensionality reduction and data representation.
\newblock {\em Neural Computation}, 15(6):1373--1396, 2003.

\bibitem{betancourt2017conceptual}
Michael Betancourt.
\newblock A conceptual introduction to hamiltonian monte carlo.
\newblock {\em arXiv preprint arXiv:1701.02434}, 2017.

\bibitem{borgwardt2006integrating}
Karsten~M. Borgwardt, Colin~S. Ong, Bernhard Sch{\"o}lkopf, SVN Vishwanathan, and Alexander~J. Smola.
\newblock Integrating structured biological data by kernel maximum mean discrepancy.
\newblock In {\em International Conference on Research in Computational Molecular Biology}, pages 78--92, 2006.

\bibitem{chiang2001tennessee}
Leo~H Chiang, Evan~L Russell, and Richard~D Braatz.
\newblock Tennessee eastman process.
\newblock In {\em Fault detection and diagnosis in industrial systems}, pages 103--112. Springer, 2001.

\bibitem{cortes2009shape}
Corinna Cortes, Mehryar Mohri, and Afshin Rostamizadeh.
\newblock Algorithms for learning kernels based on centered alignment.
\newblock {\em Journal of Machine Learning Research}, 13(28):795--828, 2012.

\bibitem{downs1993plant}
James~J. Downs and Ernest~F. Vogel.
\newblock A plant-wide industrial process control problem.
\newblock {\em Computers \& Chemical Engineering}, 17(3):245--255, 1993.

\bibitem{DUMA2026103679}
Zina-Sabrina Duma, Victoria Jorry, Tuomas Sihvonen, Satu-Pia Reinikainen, and Lassi Roininen.
\newblock Optimising kernel-based multivariate statistical process control.
\newblock {\em Journal of Process Control}, 161:103679, 2026.

\bibitem{forrest1996genetic}
Stephanie Forrest.
\newblock Genetic algorithms.
\newblock {\em ACM computing surveys (CSUR)}, 28(1):77--80, 1996.

\bibitem{gao2012implementing}
Fuchang Gao and Lixing Han.
\newblock Implementing the nelder-mead simplex algorithm with adaptive parameters.
\newblock {\em Computational Optimization and Applications}, 51(1):259--277, 2012.

\bibitem{gretton2012kernel}
Arthur Gretton, Karsten~M. Borgwardt, Malte~J. Rasch, Bernhard Sch{\"o}lkopf, and Alexander Smola.
\newblock A kernel two-sample test.
\newblock In {\em Journal of Machine Learning Research}, volume~13, pages 723--773, 2012.

\bibitem{haario2006dram}
Heikki Haario, Marko Laine, Antonietta Mira, and Eero Saksman.
\newblock Dram: efficient adaptive mcmc.
\newblock {\em Statistics and computing}, 16(4):339--354, 2006.

\bibitem{haario2001adaptive}
Heikki Haario, Eero Saksman, and Johanna Tamminen.
\newblock An adaptive metropolis algorithm.
\newblock 2001.

\bibitem{hein2005intrinsic}
Matthias Hein and Jean-Yves Audibert.
\newblock Intrinsic dimensionality estimation of submanifolds in $\mathbb{R}^d$.
\newblock {\em Proceedings of the 22nd International Conference on Machine Learning}, pages 289--296, 2005.

\bibitem{HoffmanGelman2014NUTS}
Matthew~D. Hoffman and Andrew Gelman.
\newblock The no-u-turn sampler: Adaptively setting path lengths in hamiltonian monte carlo.
\newblock {\em Journal of Machine Learning Research}, 15(1):1593--1623, 2014.

\bibitem{kavianihamedani2024new}
Hossein Kavianihamedani, Julianne~D Quinn, and Jared~D Smith.
\newblock New diagnostic assessment of mcmc algorithm effectiveness, efficiency, reliability, and controllability.
\newblock {\em IEEE Access}, 12:42385--42400, 2024.

\bibitem{KourtiMacGregor1995}
Theodora Kourti and John~F. MacGregor.
\newblock Process analysis, monitoring and diagnosis using multivariate projection methods.
\newblock {\em Chemometrics and Intelligent Laboratory Systems}, 28(1):3--21, 1995.

\bibitem{lee2020monitoring}
Wo~Jae Lee, Gamini~P Mendis, Matthew~J Triebe, and John~W Sutherland.
\newblock Monitoring of a machining process using kernel principal component analysis and kernel density estimation.
\newblock {\em Journal of Intelligent Manufacturing}, 31(5):1175--1189, 2020.

\bibitem{liu2023multivariate}
Liang Liu, Jianchang Liu, Honghai Wang, Shubin Tan, Miao Yu, and Peng Xu.
\newblock A multivariate monitoring method based on kernel principal component analysis and dual control chart.
\newblock {\em Journal of Process Control}, 127:102994, 2023.

\bibitem{meinshausen2010stability}
Nicolai Meinshausen and Peter B{\"u}hlmann.
\newblock Stability selection.
\newblock {\em Journal of the Royal Statistical Society: Series B}, 72(4):417--473, 2010.

\bibitem{moritz2016linearly}
Philipp Moritz, Robert Nishihara, and Michael Jordan.
\newblock A linearly-convergent stochastic l-bfgs algorithm.
\newblock In {\em Artificial intelligence and statistics}, pages 249--258. PMLR, 2016.

\bibitem{nickisch2008approximations}
Hannes Nickisch, Carl~Edward Rasmussen, et~al.
\newblock Approximations for binary gaussian process classification.
\newblock {\em Journal of Machine Learning Research}, 9(10):2035--2078, 2008.

\bibitem{NomikosMacGregor1995}
Panagiotis Nomikos and John~F. MacGregor.
\newblock Multivariate spc charts for monitoring batch processes.
\newblock {\em Chemical Engineering Science}, 50(15):2495--2508, 1995.

\bibitem{nomikos1995monitoring}
Petros Nomikos and John~F. MacGregor.
\newblock Monitoring batch processes using multiway principal component analysis.
\newblock {\em AIChE Journal}, 41(6):1411--1426, 1995.

\bibitem{owhadi2019kernel}
Houman Owhadi and Gene~Ryan Yoo.
\newblock Kernel flows: From learning kernels from data into the abyss.
\newblock {\em Journal of Computational Physics}, 389:22--47, 2019.

\bibitem{Panaretos2005}
John Panaretos, Stelios Psarakis, et~al.
\newblock Statistical process monitoring using multivariate methods.
\newblock {\em Quality and Reliability Engineering International}, 21(4):397--415, 2005.

\bibitem{pani2022non}
Ajaya~Kumar Pani.
\newblock Non-linear process monitoring using kernel principal component analysis: A review of the basic and modified techniques with industrial applications.
\newblock {\em Brazilian Journal of Chemical Engineering}, 39(2):327--344, 2022.

\bibitem{Pozzi2024}
Federico Pozzi et~al.
\newblock Data science for operations: Integrating statistical process control within industry 4.0 frameworks.
\newblock {\em Systems}, 12(3):100, 2024.

\bibitem{rifkin2003regularized}
Ryan Rifkin, Gregory Yeo, and Tomaso Poggio.
\newblock Regularized least squares classification.
\newblock In {\em Advances in Learning Theory: Methods, Models and Applications}, pages 131--154. IOS Press, 2003.

\bibitem{roy2007kernel}
Olivier Roy and Martin Vetterli.
\newblock The effective rank: A measure of effective dimensionality.
\newblock {\em IEEE Transactions on Signal Processing}, 55(2):451--462, 2007.

\bibitem{scholkopf2001estimating}
Bernhard Sch{\"o}lkopf, John~C. Platt, John Shawe-Taylor, Alexander~J. Smola, and Robert~C. Williamson.
\newblock Estimating the support of a high-dimensional distribution.
\newblock {\em Neural Computation}, 13(7):1443--1471, 2001.

\bibitem{scholkopf2002learning}
Bernhard Sch{\"o}lkopf and Alexander~J. Smola.
\newblock {\em Learning with Kernels: Support Vector Machines, Regularization, Optimization, and Beyond}.
\newblock MIT Press, 2002.

\bibitem{Scholkopf1998KernelPCA}
Bernhard Sch{\"o}lkopf, Alexander~J. Smola, and Klaus-Robert M{\"u}ller.
\newblock Nonlinear component analysis as a kernel eigenvalue problem.
\newblock {\em Neural Computation}, 10(5):1299--1319, 1998.

\bibitem{shawe2004kernel}
John Shawe-Taylor and Nello Cristianini.
\newblock {\em Kernel Methods for Pattern Analysis}.
\newblock Cambridge University Press, Cambridge, UK, 2004.

\bibitem{simmini2021self}
Francesco Simmini, Mirco Rampazzo, Fabio Peterle, Gian~Antonio Susto, and Alessandro Beghi.
\newblock A self-tuning kpca-based approach to fault detection in chiller systems.
\newblock {\em IEEE Transactions on Control Systems Technology}, 30(4):1359--1374, 2021.

\bibitem{tan2020}
Ruomu Tan, James~R. Ottewill, and Nina~F. Thornhill.
\newblock Monitoring statistics and tuning of kernel principal component analysis with radial basis function kernels.
\newblock {\em IEEE Access}, 8:198328--198342, 2020.

\bibitem{tax2004support}
David~MJ Tax and Robert~P.W. Duin.
\newblock Support vector data description.
\newblock {\em Machine Learning}, 54(1):45--66, 2004.

\bibitem{zelnik2004self}
Lihi Zelnik-Manor and Pietro Perona.
\newblock Self-tuning spectral clustering.
\newblock In {\em Advances in Neural Information Processing Systems}, volume~17, pages 1601--1608, 2004.

\bibitem{zhou2020multi}
Bingqian Zhou and Xingsheng Gu.
\newblock Multi-block statistics local kernel principal component analysis algorithm and its application in nonlinear process fault detection.
\newblock {\em Neurocomputing}, 376:222--231, 2020.

\end{thebibliography}

\end{document}